\pdfoutput=1
\documentclass[a4paper,fleqn,usenatbib,useAMS]{mnras}

\usepackage{graphicx}	
\usepackage{amsmath}	
\usepackage{amssymb}	
\usepackage{bm}		
\usepackage{pdflscape}	
\usepackage[english]{babel}
\usepackage{selinput}
\usepackage{multicol, lipsum, graphicx, caption}
\usepackage[toc]{appendix}
\usepackage{xcolor}

\usepackage{tabulary}
\usepackage{booktabs,longtable,tabularx}
\newcolumntype{L}{>{\raggedright\arraybackslash}X}
\usepackage{ltablex}
\usepackage{caption}

\SelectInputMappings{%
  aacute={á},
  ntilde={ñ},
  Euro={€}
}
\usepackage{babel}
\usepackage{stfloats}
\usepackage{caption}
\usepackage{subfig}

\usepackage[T1]{fontenc}
\usepackage{ae,aecompl}

\usepackage{newtxtext,newtxmath}
\usepackage{zref-xr}
\zexternaldocument{bk_compression}

\title[Compression on BOSS measurements]
{Enhancing BOSS bispectrum cosmological constraints with maximal compression}

\author[D. Gualdi et al.]
{\parbox{\textwidth}{Davide Gualdi$^{1}$\thanks{Contact e-mail: \href{davide.gualdi.14@ucl.ac.uk}{davide.gualdi.14@ucl.ac.uk}},
Héctor Gil-Marín$^{2,3}$,
Robert L. Schuhmann$^{4,5}$, 
Marc Manera$^{6,7}$, }
\newauthor{
Benjamin Joachimi$^{1}$,
Ofer Lahav$^{1}$}
\vspace{0.4cm}\
\\
\parbox{\textwidth}
{$^{1}$Department of Physics and Astronomy, University College London, Gower Street, London WC1E 6BT, UK \\
$^{2}$Sorbonne Université, Institut Lagrange de Paris (ILP), 98 bis Boulevard Arago, 75014 Paris, France\\
$^{3}$Laboratoire de Physique Nucléaire et de Hautes Energies, Université Pierre et Marie Curie, Paris, France\\
$^{4}$Institute for Astronomy, University of Edinburgh, Blackford Hill,
Edinburgh EH9 3HJ, UK\\
$^{5}$School of Physics and Astronomy, University of Manchester,
Manchester M13 9PL, UK\\
$^{6}$Institut de Física d’Altes Energies (IFAE), The Barcelona Institute of
Science and Technology, Campus UAB, 08193 Bellaterra (Barcelona) Spain\\
$^{7}$Centre for Mathematical Sciences, DAMTP, Cambridge University, Wilberforce Rd, Cambridge CB3 0WA
}}

\pubyear{2017}

\begin{document}

\label{firstpage}
\pagerange{\pageref{firstpage}--\pageref{lastpage}}
\maketitle

\begin{abstract}
We apply two compression methods to the galaxy power spectrum monopole/quadrupole and bispectrum monopole measurements from the BOSS DR12 CMASS sample. Both methods reduce the dimension of the original data-vector to the number of cosmological parameters considered, using the Karhunen-Loève algorithm with an analytic covariance model.
In the first case, we infer the posterior through MCMC sampling from the likelihood of the compressed data-vector (MC-KL). 
The second, faster option, works by first Gaussianising and then orthogonalising the parameter space before the compression; in this option (G-PCA) we only need to run a low-resolution preliminary MCMC sample for the Gaussianization to compute our posterior. 
Both compression methods accurately reproduce the posterior distributions obtained by standard MCMC sampling on the CMASS dataset for a $k$-space range of $0.03-0.12\,h/\mathrm{Mpc}$.
The compression enables us to increase the number of bispectrum measurements by a factor of $\sim 23$ over the standard binning (from 116 to 2734 triangle bins used), which is otherwise limited by the number of mock catalogues available. This reduces the $68\%$ credible intervals for the parameters $\left(b_1,b_2,f,\sigma_8\right)$ by $\left(-24.8\%,-52.8\%,-26.4\%,-21\%\right)$, respectively.
Using these methods for future redshift surveys like DESI, Euclid and PFS will drastically reduce the number of simulations needed to compute accurate covariance matrices and will facilitate tighter constraints on cosmological parameters.
\end{abstract}




\begingroup
\let\clearpage\relax
\endgroup

\begin{multicols*}{2}

\section{Introduction}
Large datasets have recently become available from current cosmological surveys (\textit{Planck}, \footnote{\url{http://sci.esa.int/planck/}} \citealp{Ade:2013zuv} ; Sloan Digital Sky Survey \footnote{\url{http://www.sdss3.org/surveys/boss.php}}, \citealp{2011AJ....142...72E}; KiDS \citealp{2013ExA....35...25D}; DES, \citealp{2016MNRAS.460.1270D} \footnote{\url{https://www.darkenergysurvey.org}})
and even larger ones will be provided in future by DESI\footnote{\url{http://desi.lbl.gov}}, \citet{Levi:2013gra}; Euclid \footnote{\url{http://sci.esa.int/euclid/}}, \citet{2011arXiv1110.3193L}; PFS \footnote{\url{http://pfs.ipmu.jp}}, \citet{2014PASJ...66R...1T} and the LSST\footnote{\url{https://www.lsst.org/}}, \citet{2009arXiv0912.0201L}.
In order to exploit their full potential, is desirable to go beyond standard two-points statistics (2pt). 

Three-points statistcs (3pt) are a complementary probe that is possible to investigate both in configuration and Fourier space and have been used extensively in galaxy clustering analyses (\citealp{1977ApJ...217..385G}, \citealp{1984ApJ...279..499F}, \citealp{1993ApJ...413..447F}, \citealp{1994ApJ...425..392F}, \citealp{Matarrese:1997sk}, \citealp{Verde:1998zr}, \citealp{Heavens:1998es}, \citealp{Scoccimarro:1997st}, \citealp{2000ApJ...544..597S}, \citealp{2006PhRvD..74b3522S}).
Deviations from General Relativity \citep{PhysRevD.79.103506,Bernardeau:2011sf,2011JCAP...11..019G} and primordial non-Gaussianities \citep{1994ApJ...429...36F,1994ApJ...430..447G,doi:10.1046/j.1365-8711.2000.03191.x, 2010AdAst2010E..73L, 2016JCAP...06..014T} have been investigated using 3pt statistics.
Their potential in lifting degeneracies present at 2pt level has been shown by the most recent measurement on the BOSS dataset, for the bispectrum by \citet{2017MNRAS.465.1757G} and for the 3pt correlation function by \citet{2017MNRAS.468.1070S}. Baryonic acoustic oscillations (BAO) have also been measured using the 3pt correlation function by \citet{2017MNRAS.469.1738S} and detected using the bispectrum by \citet{2017arXiv171204970P}. 

Recently, 3pt statistics have been studied in the case of 21cm emission lines by \citet{2018arXiv180202578H}.
For what concerns weak lensing, its effect on 3pt galaxy clustering have been studied by \citet{PhysRevD.78.043513}. Moreover the weak lensing bispectrum has been object of several studies in recent years (\citealp{2004MNRAS.348..897T,2009A&A...508.1193J,2013MNRAS.429..344K,2013arXiv1306.4684K}). The skewness of mass aperture statistic was considered by \citet{2004MNRAS.352..338J} while the 3pt correlation function of cosmic shear was analysed by \citet{2005A&A...431....9S,2005A&A...442...69K}.
Higher order statistics like the bispectrum via gravitational lensing have been investigated also by \citet{2013MNRAS.430.2476S,2014MNRAS.441.2725F,2015MNRAS.449.1505S,2017JCAP...12..043P}.

Besides being computationally more expensive than 2pt statistics, 3pt statistics present the drawback to be described by very large data-vectors, which in turn require a high number of simulations to accurately estimate their covariance matrix \citep{2007A&A...464..399H}.
In \citet{2018MNRAS.tmp..252G}, Paper I from now on, we presented two methods to compress the redshift-space galaxy bispectrum, namely MC-KL (Markov chain Monte Carlo sampling + Karhunen-Loève compression) and PCA + KL (principal component analysis transformation + Karhunen-Loève compression). MC-KL consists in sampling via MCMC the compressed data-vector's likelihood. PCA + KL reconstructs the multidimensional physical posterior distribution from the 1D posterior of orthogonalised parameters obtained by diagonalising the Fisher information matrix. Modifications/improvements of the Karhunen-Loève algorithm were introduced also by \citet{2000MNRAS.317..965H} and recently by \citet{2017MNRAS.472.4244H,2018MNRAS.476L..60A,2018MNRAS.477.2514A} also with the target of data compression.

In this work we apply our compression methods to both the power spectrum monopole/quadrupole and to the bispectrum monopole measurements from the CMASS sample of BOSS DR12.
While the MC-KL is more flexible than the PCA + KL method since doesn't require the multidimensional Gaussian posterior assumption, the PCA + KL is much faster in terms of computational time and requires far fewer computational resources (it can be run on standard laptop). We compare both methods and test their convergence in terms of deriving equivalent posterior distributions.

In order to make the PCA + KL method applicable also to parameter spaces with strong degeneracies, for which the posterior Gaussianity approximation is no longer valid, we introduce a pre-Gaussianisation step based on the algorithm developed by \citet{2016MNRAS.459.1916S}.

We measure the bispectrum monopole using the same code used for the BOSS DR12 analysis done by \citet{2017MNRAS.465.1757G}. We vary the size of the triangle vectors by changing the bin size $\Delta k$ for $k$, which returns different number of triangular shapes given the minimum and maximum scales. For the same number of triangle bins the compression returns posterior distributions slightly larger than the MCMC counterparts. However, when compressing a much larger number of triangle bins (which cannot be done for the MCMC on the full data-vector because of the limited number of mocks available constraint), the posterior distribution becomes more Gaussian and narrow. It eventually returns tighter constraints than the ones obtained by the standard analysis.

In Sec. \ref{sec:datav_cov} we present the analytical model used for the data-vector considered and the analytical expression of the covariance matrix used to derive the weights for the compression.
In Sec. \ref{sec:data_mocks_analysis} we describe the data set and the galaxy mocks used to estimate the covariance matrix together with the settings of our analysis. In Sec. \ref{sec:comp_meth} we recap the compression methods applied including the Gaussianisation extension for the original PCA + KL method. We report the performance of the compression methods compared to the MCMC sampling for the cases in which it is possible to run it on the full data-vector in Sec. \ref{sec:recover_posterior}. We describe the gain in parameter constraints as a function of the number of triangle bins used in the bispectrum monopole data-vector component in Sec. \ref{sec:info_tr_num}. We test the flexibility and accuracy of the compression methods presented in Sec. \ref{sec:consistency_check}.
Finally we conclude summarising our results in Sec. \ref{sec:conclusions}. In Appendix \ref{sec:est_def} we report the full derivation of all the analytic expressions used in the analysis. In Appendix \ref{sec:validation} additional validation tests are presented.

\section{data-vector and Covariance Matrix}
\label{sec:datav_cov}

In order to measure the power spectrum and bispectrum from the data and the mocks catalogues we use the estimators  described in  \cite{2016MNRAS.460.4188G,2016MNRAS.460.4210G}. These are based on the weighted field of density fluctuations \citep{1994ApJ...426...23F}:

\begin{eqnarray}
F_{\lambda}(\bm{r})\,=\,\dfrac{w_{\mathrm{FKP}}(\bm{r})}{I^{1/2}_{\lambda}}[w_{\mathrm{c}}(\bm{r})n(\bm{r}) - \alpha n_{\mathrm{syn}}(\bm{r})],
\end{eqnarray}
    
\noindent where $w_{\mathrm{c}}$ is the weight taking into account all the measurement systematics (redshift failure, fiber collision, target density variations), $w_{\mathrm{FKP}}$ (Feldman, Kaiser and Peacock) ensures the condition of minimum variance, $n$ is the observed number density of galaxies, $n_{\mathrm{syn}}$ is the number density of objects in a synthetic catalogue and $I_{\lambda}$ is the normalisation of the amplitude of the observed power ($\lambda=2,3$ for power spectrum and bispectrum, respectively). $\alpha$ is the ratio between weighted number of observed galaxies over the weighted number of objects in the synthetic catalogues.

\subsection{Power spectrum monopole and quadrupole}
The redshift-space galaxy power spectrum model adopted in this work is a linear one including redshift-space distortions (RSD) plus a damping function taking into account the Finger-of-God (FoG) effect:

\begin{eqnarray}
\label{pk_esp}
\mathrm{P}_{\mathrm{g}}^{\mathrm{s}}\left(k, \mu\right) 
=\mathrm{D}^{\mathrm{P}}_{\mathrm{FoG}}\left(k, \mu, \sigma^{\mathrm{P}}_{\mathrm{FoG}}[z]\right) \mathrm{Z}_{1}^{\mathrm{s}}\left(\bm{k}\right)^2\mathrm{P}_{\mathrm{m}}^{\mathrm{lin.}}\left(k\right)\,,
 \end{eqnarray}

\noindent where $k$ is the module of the wave vector $\bm{k}$ and $\mu$ is the cosine of the angle between the wave vector and the line of sight. The standard redshift-space distortion kernels $\mathrm{Z}^{\mathrm{s}}_i$ are reported in the Appendix of \citet{2014JCAP...12..029G} together with the FoG damping function expression. $\sigma^{\mathrm{P}}_{\mathrm{FoG}}[z]$ is the FoG free parameter for the power spectrum. 
For the range of scales considered in this work the linear RSD model has been tested on N-body simulations and proved to be a good approximation (\citealp{2010PhRvD..82f3522T}, Figure 2). The redshift-space galaxy power spectrum can be expanded in terms of Legendre polynomials using its dependence on $\mu$:

\begin{eqnarray}
 \mathrm{P}_{\mathrm{g}}^{\mathrm{s}}\left(k, \mu\right) = \sum^{\infty}_{\ell=0} \mathrm{P}_{\mathrm{g}}^{(\ell)}\left(k\right)L_{\ell}\left(\mu\right)\,,
\end{eqnarray}

\noindent where $L_{\ell}\left(\mu\right)$ is the $\ell$-order Legendre polynomial. Almost all the signal is contained in the first two even multipoles, the monopole and the quadrupole ($\ell=0,2$). These can be found by inverting the above expression:

\begin{eqnarray}
\label{pk_02_esp}
 \mathrm{P}_{\mathrm{g}}^{(\ell)}\left(k\right) = \dfrac{2\ell+ 1}{2}\int^{+1}_{-1}d\mu\,\mathrm{P}_{\mathrm{g}}^{\mathrm{s}}\left(k, \mu\right)L_{\ell}\left(\mu\right)\,.
\end{eqnarray}

\subsection{Analytical expression for $\mathrm{P}^{(0,2)}_{\mathrm{g}}$ covariance matrices}
Defining an estimator as in Appendix \ref{sec:estimators}, it is possible to derive the expression for the Gaussian term of the power spectrum monopole and quadrupole covariance matrices (Appendix \ref{sec:cov_pp}):

\begin{align}
\label{eq:cov_p02}
&\mathrm{C}_{\mathrm{G}}^{\mathrm{P}^{(\ell)}_{\mathrm{g}}\mathrm{P}^{(\ell)}_{\mathrm{g}}}\left(k_1;k_2\right) = 
\left(\dfrac{2\ell+1}{2}\right)^2\dfrac{2\delta^{\mathrm{K}}_{12}}{N_{\mathrm{p}}\left(k_1\right)}\quad\mathrm{P}_{\mathrm{g}}^{(\ell)}\left(k_1\right)^2 ,
\end{align}

\noindent where $\delta^{\mathrm{K}}_{12}$ is the Kronecker delta between $k_1$ and $k_2$, while $N_{\mathrm{p}}(k_1)$ is the number of pairs of grid points inside the estimator integration volume in Fourier space  $V_k = 4\pi k^2\Delta k$ \citep{1998ApJ...496..586S} and it is proportional to an effective survey volume $V_{\mathrm{e}}$. The $V_{\mathrm{e}}$ normalisation is used to obtain a closer match between the analytic and mocks covariance matrices (please refer to Eqs. \ref{eq:n_pairs} and \ref{eq:cov_pp} for more details). We set the cross covariance between power spectrum monopole and quadrupole to zero since it is negligible w.r.t. the other terms, as can be seen from Figure 3 in \citet{2017MNRAS.465.1757G}.

\subsection{Bispectrum monopole}
For the redshift-space galaxy bispectrum we adopt the effective model presented in \citet{2014JCAP...12..029G}, which modifies the redshift-space distortion kernels derived from perturbations theory in order to better fit the data at non-linear scales (see the Appendix of the paper above for the full expressions). This effective model includes 18 parameters which have been calibrated using simulations \citep{2012JCAP...02..047G,2014JCAP...12..029G}. The model has been applied to both BOSS DR11 and DR12 data-sets \citep{2015MNRAS.451..539G,2017MNRAS.465.1757G}.
The tree level has also been corrected to take into account the Finger-of-God damping effect:

\begin{align}
&\mathrm{B}_{\mathrm{g}}^{\mathrm{s}}\left(\bm{k}_1,\bm{k}_2,\bm{k}_3\right) 
=
\mathrm{D}^{\mathrm{B}}_{\mathrm{FoG}}\left(\bm{k}_1,\bm{k}_2,\bm{k}_3,\sigma^{\mathrm{B}}_{\mathrm{FoG}}[z]\right)
\notag \\
&\times
\left[\mathrm{Z}_{1}^{\mathrm{s}}\left(\bm{k}_1\right)\mathrm{Z}_{1}^{\mathrm{s}}\left(\bm{k}_2\right)
\mathrm{Z}_{2,\mathrm{eff.}}^{\mathrm{s}}\left[\bm{k}_1,\bm{k}_2\right]
\mathrm{P}_{\mathrm{m}}^{\mathrm{lin.}}\left(k_1\right)\mathrm{P}_{\mathrm{m}}^{\mathrm{lin.}}\left(k_2\right) + \mathrm{cyc.}\right]\,,
\end{align}

\noindent where $\sigma^{\mathrm{B}}_{\mathrm{FoG}}[z]$ is the FoG free parameter for the bispectrum. The monopole of the bispectrum corresponds to the average of all the possible orientations of a certain triangle, given by three wave-vectors' moduli, with respect to the line of sight. It can therefore be obtained by integrating over two angular coordinates:

\begin{align}
\label{bk_0_esp}
\mathrm{B}^{(0)}_{\mathrm{g}}\left(k_1,k_2,k_3\right)
&= \dfrac{1}{4}\int^{1}_{-1} d\mu_1\int^{1}_{-1}d\mu_2 \,\mathrm{B}^{\mathrm{s}}_{\mathrm{g}}\left(\bm{k}_1,\bm{k}_2,\bm{k}_3\right)
\notag \\
&=\dfrac{1}{4\pi}\int^{1}_{-1} d\mu_1\int^{2\pi}_{0}d\phi \, \mathrm{B}^{\mathrm{s}}_{\mathrm{g}}\left(\bm{k}_1,\bm{k}_2,\bm{k}_3\right)
\,,
\end{align}

\noindent where $\mu_i$ is the cosine of the angle between the $\bm{k}_i$ vector and the line of sight. The angle $\phi$ is defined as $\mu_2\equiv\mu_1x_{12} - \sqrt{1-\mu_1^2}\sqrt{1-x_{12}^2}\cos{\phi}$ and where $x_{12}$ is the cosine of the angle between $\bm{k}_1$ and $\bm{k}_2$. More details are given in Appendix \ref{sec:est_def}.

\subsection{Analytical expression for $\mathrm{B}^{(0)}_{\mathrm{g}}$ covariance matrix}

In order to apply the compression methods presented in Paper I we need an analytical expression for the bispectrum monopole covariance matrix. 
This allows us to compress a data-vector with an arbitrarily large number of triangle bins, which on the contrary wouldn't be possible using a covariance matrix estimated from the galaxy mock catalogues. That is because in order to obtain an accurate numerical estimate of the covariance matrix, the number of simulations used must be much greater than the data-vector's dimension \citep{2007A&A...464..399H,2014MNRAS.439.2531P}.

As it has been shown in Paper I, compressing the power spectrum together with the bispectrum, or leaving it uncompressed, does not make any substantial difference in terms of recovered parameter constraints. However, it makes a huge difference in terms of complexity of the covariance matrix that one has to model analytically in order to compress the data-vector. Compressing the power spectrum as well (monopole and quadrupole) also requires modelling their covariance matrices together with the cross-covariance with the bispectrum monopole. 
Leaving them uncompressed just requires to model the bispectrum monopole covariance matrix.

The covariance terms for the bispectrum monopole below reported are original of this work. Expressions for the matter bispectrum were derived also by \citet{1998ApJ...496..586S,2006PhRvD..74b3522S,2017PhRvD..96b3528C}, however in order to compute covariance matrix we proceed similarly to what done in \citet{2013MNRAS.429..344K}. 

The expression for the Gaussian term of $\mathrm{C}^{\mathrm{B}^0_{\mathrm{g}}\mathrm{B}^0_{\mathrm{g}}}$ is derived in Appendix \ref{sec:cov_bb_g} and reads:

\begin{align}
\label{eq:cov_b0}
&\mathrm{C}_{\mathrm{G}}^{\mathrm{B}^0_{\mathrm{g}}\mathrm{B}^0_{\mathrm{g}}}\left(k_1,k_2,k_3;k_4,k_5,k_6\right) =
\notag \\
&=
\dfrac{\mathrm{D}_{123456}}{16\pi^2}\dfrac{V_{\mathrm{e}}}{N_{\mathrm{t}}\left(k_1,k_2,k_3\right)}
\mathrm{P}_{\mathrm{g}}^{(0)}\left(k_1\right)\mathrm{P}_{\mathrm{g}}^{(0)}\left(k_2\right)\mathrm{P}_{\mathrm{g}}^{(0)}\left(k_3\right) ,
\end{align}

\noindent where $\mathrm{D}_{123456}$ stands for all the possible permutations for which each side of the first triangle is equal to a side of the second one; it has the values $(6,2,1)$ respectively for equilateral, isosceles and scalene triangles.  $N_{\mathrm{t}}\left(k_1,k_2,k_3\right)$ is the number of independent triplets of grid points in the integration volume in Fourier space $V_{k_{123}}\simeq 8\pi^2 k_1k_2k_3\Delta k_1\Delta k_2\Delta k_3$ . For the values of the effective survey volume and the average galaxy density number used in computing the analytical covariance matrix, we adopt the values $V_{\mathrm{e}} = 2.43 \times 10^9 \,\mathrm{Mpc}^3$ and $\bar{n}_{\mathrm{g}} =  1.14 \times \, 10^{-4}\mathrm{Mpc}^{-3}$ used by \citet{2017MNRAS.468.1070S} for both power spectrum monopole/quadrupole and bispectrum monopole analytical covariance matrices. In practice we use the analytic expression of the covariance matrix only to determine the weights for the compression. Since all the terms considered scale as $V_{\mathrm{e}}^{-1}$ the effective volume acts only as a scaling factor not affecting the compression performance.

In order to describe the correlation between different triangle bins in our analytical model of the covariance matrix, we include also a non-Gaussian term of the bispectrum monopole covariance matrix. In the expansion of the bispectrum covariance matrix presented in the Appendix of Paper I, for the bispectrum monopole this corresponds to a term proportional to the product of two bispectra monopoles as shown in Appendix \ref{sec:cov_bb_ng}:

\begin{align}
\label{eq:ng_bb}
  &\mathrm{C}_{\mathrm{NG}}^{\mathrm{B}^0_{\mathrm{g}}\mathrm{B}^0_{\mathrm{g}}}\left(k_1,k_2,k_3;k_4,k_5,k_6\right) =
\notag \\
&=  
\dfrac{\delta^{\mathrm{K}}_{34}}{16\pi^2}\dfrac{k_f^{3}}{4\pi k_3^2\Delta k_3}
\mathrm{B}_{\mathrm{g}}^{(0)}\left(k_1,k_2,k_3\right)\mathrm{B}_{\mathrm{g}}^{(0)}\left(k_3,k_5,k_6\right)
\quad + 8 \quad\mathrm{perm.}
\end{align}

\noindent It is important to include a term modelling the correlation between different triangle bins since the number of possible configurations increases very quickly as the bin size decreases. 
For simplicity we only used this term to model the correlation between different triangle bins. However it is important to notice that a better approximation of the analytical covariance matrix can be obtained by including the expressions corresponding to all the terms present in the expansion given in the Appendix of Paper I.

We do not include a corresponding non-Gaussian term into the power spectrum monopole and quadrupole covariances, since the number of data points considered is relatively low, thus the separation between the $k$ modules values is more than sufficient to assume that the correlation between two different modes $k_i$ and $k_j$ is negligible with respect to their variance (approximated by the Gaussian term on the diagonal of the covariance matrix). 
From Figure 3 in \citet{2017MNRAS.465.1757G} it can be seen that the cross-variance between different data-points for the monopole and quadrupole of the power spectrum is much weaker than their variance.

\subsection{Analytical expression for $\left[\mathrm{P}^{(0,2)}_{\mathrm{g}},\mathrm{B}^{(0)}_{\mathrm{g}}\right]$ cross-covariance matrix}
Finally we also model the cross-covariance between power spectrum multipoles and bispectrum monopole as described in Appendix \ref{sec:cov_cross}:

\begin{align}
\label{eq:ct_pb}
&\mathrm{C}^{\mathrm{P}^{(\ell)}_{\mathrm{g}}\mathrm{B}^0_{\mathrm{g}}}\left(k_1;k_2,k_3,k_4\right) = 
\notag \\
&=
\dfrac{1}{2\pi}\left(\dfrac{2\ell+1}{2}\right)\dfrac{\delta^{\mathrm{K}}_{12}}{N_{\mathrm{p}}\left(k_2\right)}
\mathrm{P}^{(\ell)}_{\mathrm{g}}\left(k_2\right)\mathrm{B}^{(0)}_{\mathrm{g}}\left(k_2,k_3,k_4\right) \quad + 2 \quad\mathrm{perm.}
.
\end{align}

\noindent As done in Paper I, we made the assumption that the shot noise is well approximated by a Gaussian distribution (which is reasonable if the galaxy number density is fairly high). Therefore, we just modify the galaxy power spectrum expressions by adding a $\bar{n}_{\mathrm{g}}^{-1}$ term.
We did not take into account the effect of the survey geometry in the theoretical covariance matrix expression, which would affect the large scales inducing an extra correlation among the modes. We leave the inclusion of this correction for future work. Please refer to \citet{2017MNRAS.472.4935H} for a more detailed study on how to include this effect in the covariance matrix.

\section{DATA, MOCKS and ANALYSIS}
\label{sec:data_mocks_analysis}
\subsection{DR12 BOSS data and mocks catalogues}
\label{sec:data_and_mocks}
In this paper we use the CMASS galaxy sample ($0.43\leq z \leq 0.70$) of the Baryon Oscillation Spectroscopic Survey (BOSS \citealp{2013AJ....145...10D}) which is part of the Sloan Digital Sky Survey III \citep{2011AJ....142...72E}. In the final data release DR12 the CMASS sample contains the spectroscopic redshift of 777202 galaxies (see \citealt{2017MNRAS.465.1757G} and \citealp{2017MNRAS.470.2617A} for more details).

In order to accurately numerically estimate the covariance matrix it is necessary to employ a large suite of mock galaxy catalogues. These are different realizations of the same region of the Universe based on methods such as second-order Lagrangian perturbation theory (2LPT \citealt{2002MNRAS.329..629S,2013MNRAS.428.1036M}) or augmented Lagrangian perturbation theory (ALPT) as described in \citet{2013MNRAS.435L..78K}. By measuring the data-vector of interest on each one of these catalogues we can numerically estimate the covariance matrix which will be used in the likelihood evaluation. In this work we use subsets of the 2048 realisations of the MultiDark Patchy BOSS DR12 mocks by \citet{2016MNRAS.456.4156K}. This set of mocks has been run using the underlying cosmology: $\Omega_{\Lambda} = 0.693$, $\Omega_{\mathrm{m}}(z=0) = 0.307$, $\Omega_{\mathrm{b}}(z=0) = 0.048$, $\sigma_8(z=0)=0.829$, $n_{\mathrm{s}} = 0.96$, $h_0 = 0.678$.

\subsection{Analysis settings}
\label{sec:analysis_settings}
For the power spectrum monopole and quadrupole the bin size was fixed to $\Delta k = 0.01 h/\mathrm{Mpc}$.
We measured the bispectrum monopole from both data and mocks using different multiples of the fundamental frequency defined as $k_f^3 = \frac{(2\pi)^3}{V_{\mathrm{s}}}$ where $V_{\mathrm{s}}$ is the survey volume which in this case was the cubic box volume $V_{\mathrm{s}} = L_{\mathrm{b}}^3 = (3500\, \mathrm{Mpc}/h)^3$ used to analyse the galaxy mocks. In particular, the considered bin sizes for the bispectrum are $\Delta k =(6,5,4,2)\, \times\, k_f$ respectively, corresponding to 116, 195, 404 and 2734 triangle bins used between $0.02<k_i\, [h/\mathrm{Mpc}] < 0.12$. For every $\Delta k_i$ bin size all the measured triangle bins bispectra, which depends on the chosen bin size, are regrouped in the number of triangle bins above specified. The largest bin size $\Delta k =6\, \times\, k_f$ corresponds to the one used in the BOSS collaboration analysis done by \citet{2017MNRAS.465.1757G}. For the $k$-range considered in the BOSS analysis the $\Delta k_6$ ($\Delta k=6\times k_f$) binning case corresponded to 825 triangle bins (triplets of wave-vector's modulus) while $\Delta k_2$ would have corresponded to more than $\sim7000$ triangle bins.

In all the parameter estimation analyses that we are going to perform, we use the covariance matrix derived from the galaxy catalogues described above (see Sec. \ref{sec:data_and_mocks}). In particular, we use 1400 mocks to estimate the covariance matrix when running the MCMC sampling on the full data-vector. We use 700 when the analysis is performed using the compressed data-vector.

The largest scales considered in this work are $k_{\mathrm{min}} = 0.03\, h/\mathrm{Mpc}$ for both power spectrum monopole and quadrupole and $k_{\mathrm{min}} = 0.02\, h/\mathrm{Mpc}$ for the bispectrum monopole. 
The choice of $k_{\mathrm{min}} = 0.03\, h/\mathrm{Mpc}$ reduces the impact of the large scale systematic errors on the analysis \citep{2016MNRAS.460.4188G}. In the case of the bispectrum, since the model is more accurate and very similar to the one that it has been already applied and tested on data in the BOSS analysis, we preferred to use a lower $k_{\mathrm{min}}$ in order to be able to use a wider range of triangle bins.

The smallest scales considered are $k_{\mathrm{max}} = 0.09\, h/\mathrm{Mpc}$ and $k_{\mathrm{max}} = 0.12\, h/\mathrm{Mpc}$ for power spectrum (monopole and quadrupole) and bispectrum monopole respectively. The lower $k_{\mathrm{max}}$ used for the power spectrum is due to the fact that we did not include 1-loop corrections for it, hence we consider only scales belonging to the quasi-linear regime. We chose a higher $k_{\mathrm{max}}$ for the bispectrum since we implemented the effective model developed by \citet{2014JCAP...12..029G} which works up to non-linear scales.

The fiducial cosmology chosen for the analysis corresponds to a flat-$\Lambda$CDM model close to the one reported in
\cite{2016A&A...594A..13P}. In particular, we set $\Omega_{\mathrm{m}}(z=0) = 0.31$, $\Omega_{\mathrm{b}}(z=0) = 0.049$, $A_{\mathrm{\mathrm{s}}}=2.21\times 10^{-9}$, $n_{\mathrm{s}} = 0.9624$, $h_0 = 0.6711$.
In order to compute the covariance terms and the derivatives of the model necessary for the compression, we fix the fiducial value of the galaxy bias model parameters, the growth rate and the amplitude of dark matter fluctuations to the ones obtained by running a preliminary low-resolution MCMC ($b_1 =2.5478$, $b_2=1.2127$, $f =0.7202$, $\sigma_8 = 0.4722$). The Finger-of-God parameters for both power spectrum and bispectrum $\sigma_{\mathrm{FoG}}^{\mathrm{B}} $ and $\sigma_{\mathrm{FoG}}^{\mathrm{P}}$ have been set to zero after checking that for the range of scales considered (quasi-linear regime) they were compatible with zero.
In Section \ref{sec:consistency_check} we check that the choice of fiducial parmeters used to compute the derivatives of the mean of the data-vector and the analytical covariance matrix does not significantly influence the results of the compression.

\section{Compression Methods}
\label{sec:comp_meth}
In Paper I we presented two compression methods and applied them to the galaxy bispectrum and power spectrum: MC-KL and PCA + KL. Both methods rely on the Karhunen-Loève algorithm (KL) applied for the first time for multi-parameter inference in cosmology by \cite{1997ApJ...480...22T}. Using this KL compression it is possible to shrink an arbitrarily large data-vector $\bm{x}$ to a compressed one $\bm{y}$ having dimension equal to the number of model parameters considered preserving Fisher information. This is obtained by deriving a set of weights for the full data-vector for each model parameter. Taking the scalar product between the weighting vectors and the original full data-vector $\bm{x}$ gives the elements $y_i$ of the compressed data-vector. Here we report only the main equations, please refer to Paper I for more details. The weighting vector for each parameter $\theta_i$ is given by:

\begin{eqnarray}
\label{eq:weights}
\bm{b} = \mathrm{\textbf{Cov}}_{\bm{x}}^{-1} \langle\bm{x}\rangle_{,i} \,,
\end{eqnarray}

\noindent where $\mathrm{\textbf{Cov}}^{-1}$ is the inverse of the original full data-vector covariance matrix and $\langle\bm{x}\rangle_{,i}$ is the derivative with respect to the model parameter $\theta_i$ of the mean of the modelled data-vector $\bm{x}$, computed at a fiducial parameter vector $\bm{\theta}_{\mathrm{fid.}}$ . In our case the fiducial values are reported in Section \ref{sec:analysis_settings}. Therefore, the elements of the compressed data-vector $\bm{y}$ are given by:

\begin{eqnarray}
y_i\, = \,\langle\bm{x}\rangle_{,i}^\intercal\,\mathrm{\textbf{Cov}}_{\bm{x}}^{-1}\,\bm{x} \,\equiv\, \bm{b}^\intercal\,\bm{x}.
\end{eqnarray}

\noindent In the MC-KL method a MCMC sampling algorithm using $\bm{y}$ as data-vector is ran after compression. An estimate of the compressed covariance matrix from the mock catalogues can be obtained as shown in the Appendix of Paper I:

\begin{eqnarray}
\label{comp_cov}
\mathrm{\textbf{Cov}}_{\bm{y},ij}\,=\,\mathrm{\textbf{Cov}}\left[y_i,y_j\right]
\,=\,
\bm{b}_i^\intercal \cdot \mathrm{\textbf{Cov}}_{\bm{x}} \cdot \bm{b}_j\,,
\end{eqnarray}

\noindent where $\mathrm{\textbf{Cov}}_{\bm{x}}$ is the original covariance matrix.

\begin{figure*}%
    \centering
    \subfloat[consistency check]{{\includegraphics[width=0.5\textwidth]
    {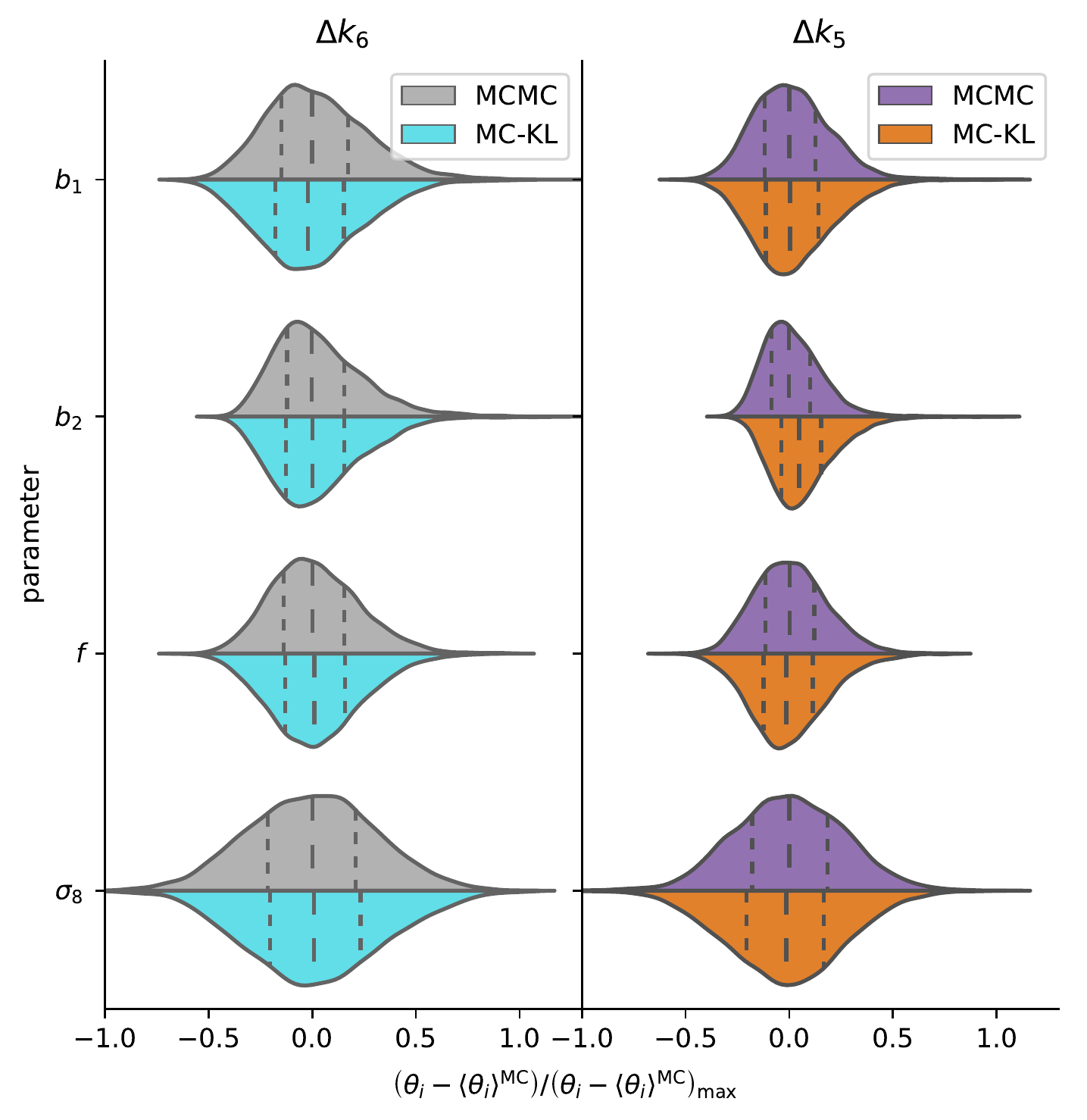}
    }}%
    \subfloat[constraints improvement]{{\includegraphics[width=0.5\textwidth]
    {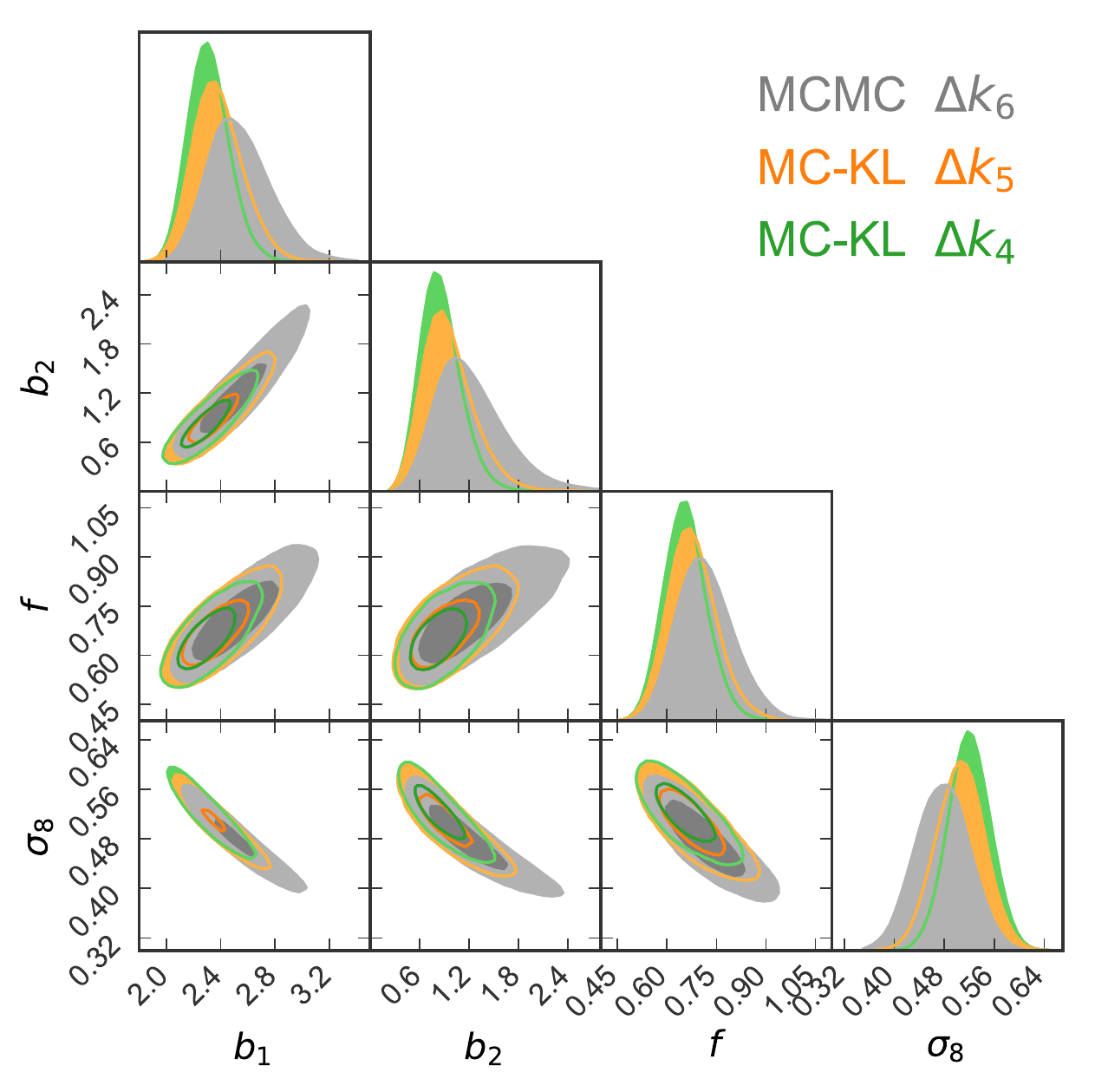}
    }}%
    \caption{ Joint data-vector $\left[\mathrm{P}^{(0)}_{\mathrm{g}},\mathrm{P}^{(2)}_{\mathrm{g}},\mathrm{B}^{(0)}_{\mathrm{g}}\right]$ posteriors: MC-KL four-parameter case. \newline
    \textbf{a)}
    the violin plots show for two test cases ($\Delta k_6$ and $\Delta k_5$ binning) the comparison between the 1D posterior densities obtained via MCMC and MC-KL for all parameters. The vertical lines represent the $25\%$, $50\%$ and $75\%$ quartiles. All distributions have been centered by subtracting the mean value obtained from the MCMC analysis and they have been normalised by dividing by the maximum difference between the parameter value of each sample and the mean of the distribution. Even if the 1D distributions are not Gaussian, the agreement between MCMC and MC-KL results is qualitatively good. For a quantitative comparison see Table \ref{tab:4pars_consistency} and additionally Figure \ref{fig:mcmc_vs_MC-KL_vs_pc_check} and \ref{fig:mcmc_vs_MC-KL_vs_pc_check2} in Appendix \ref{sec:validation}. \newline
    \textbf{b)}
    the 2D $68\%$ and $95\%$ credible regions are shown in order to highlight the improved constraints and reduced parameter degeneracies obtained by employing a higher number of triangle bins in the data-vector thanks to the compression with respect to the standard MCMC for the full data  vector. In particular, the grey contours correspond to the standard binning $\Delta k_6$ used to run the MCMC for the full data-vector. The orange and green
    contours correspond to the distributions for the compressed data-vector for the binnings $\Delta k_5$ and $\Delta k_4$
    (which corresponds to $\mathrm{N}_{\mathrm{triangles}} = 195,\,404$, the number of triangle bins increases as the $k$-bin size approaches the fundamental frequency). See also Table \ref{tab:4pars_improvement}.}
    
    \label{fig:mcmc_vs_MC-KL_check_4pars}
\end{figure*}


\begin{figure*}%
    \centering
    \subfloat[consistency check]{{\includegraphics[width=0.5\textwidth]
    {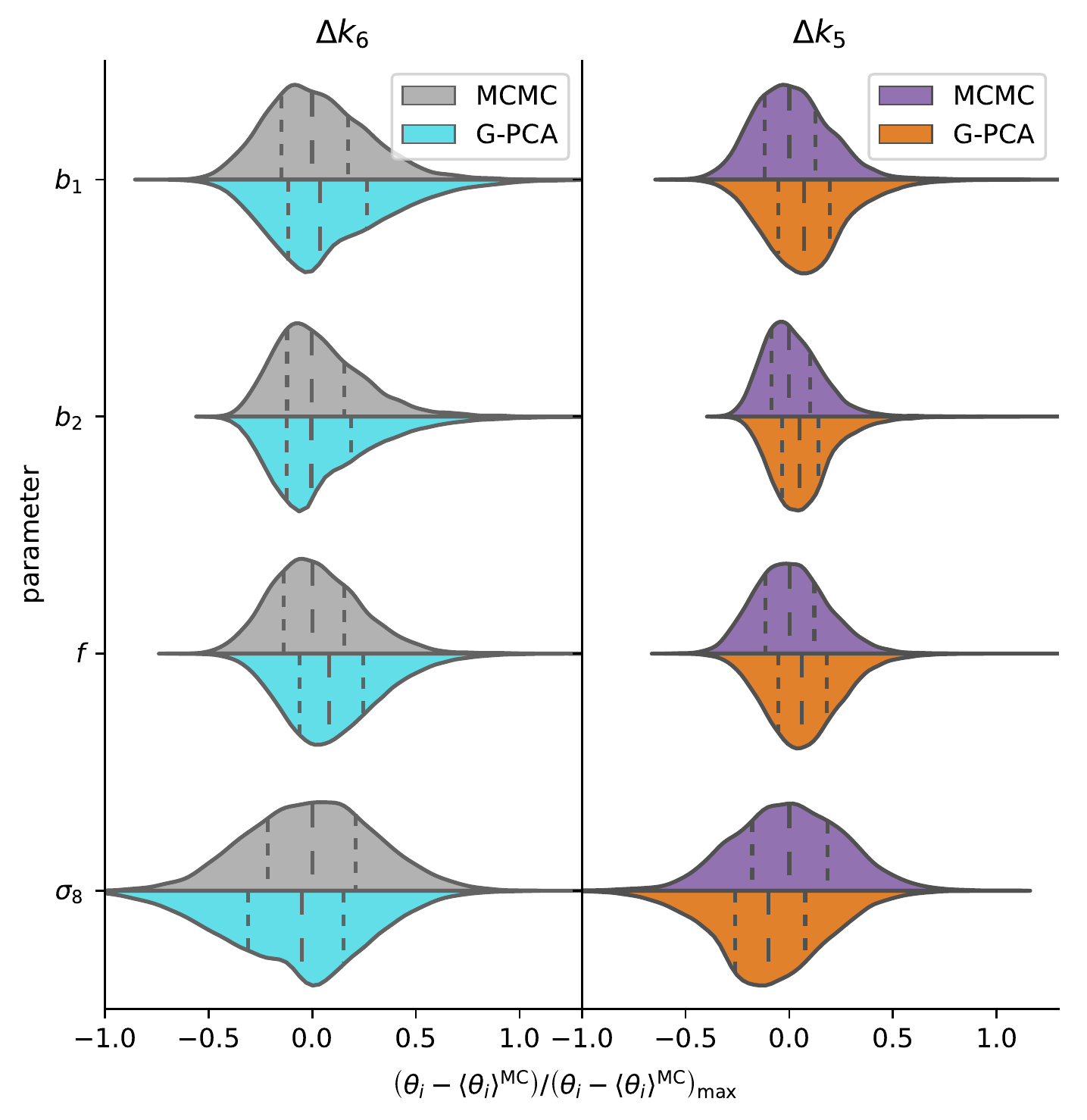}
    }}%
    \subfloat[constraints improvements]{{\includegraphics[width=0.5\textwidth]
    {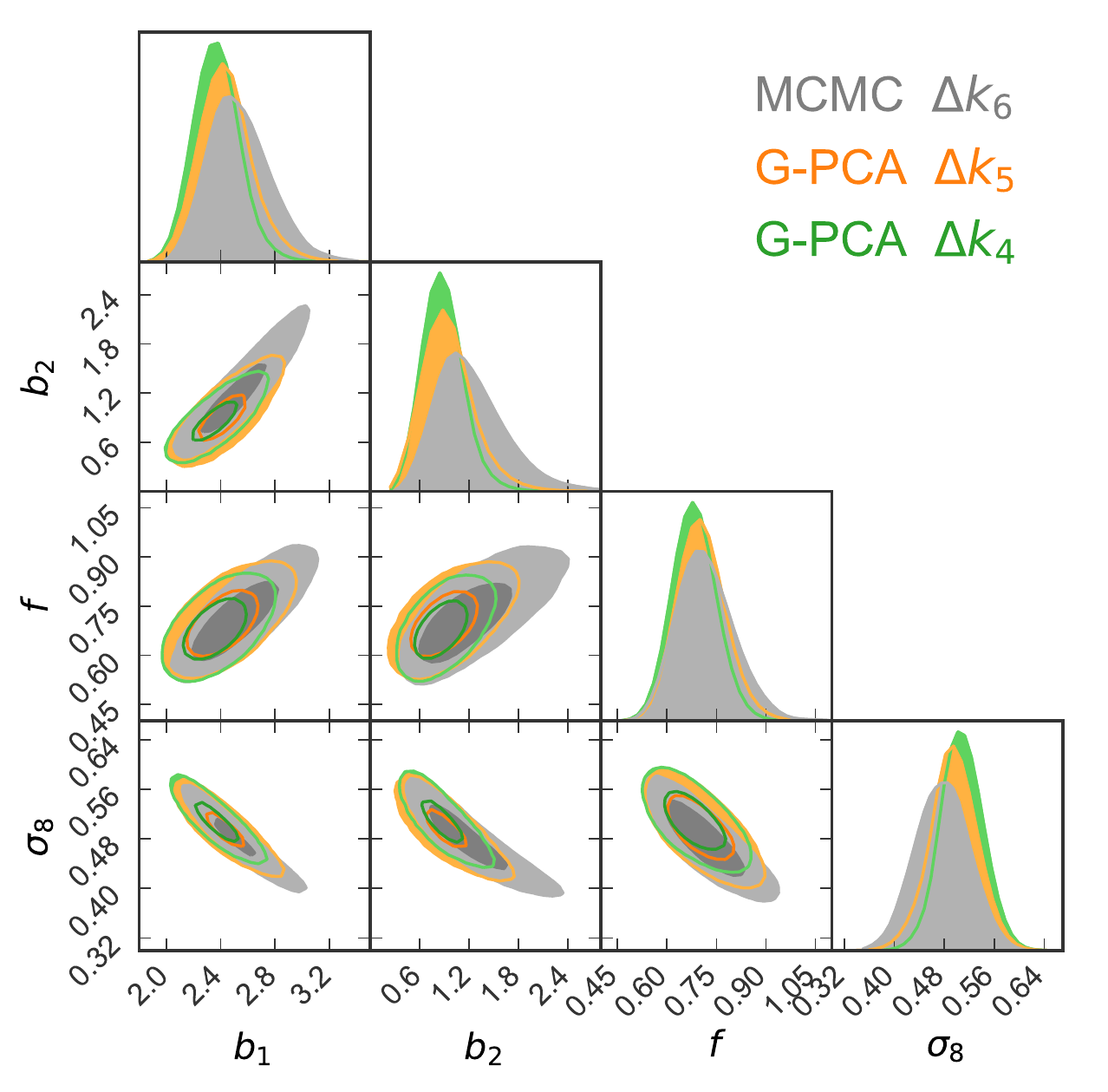}
    }}%
    \caption{Joint data-vector $\left[\mathrm{P}^{(0)}_{\mathrm{g}},\mathrm{P}^{(2)}_{\mathrm{g}},\mathrm{B}^{(0)}_{\mathrm{g}}\right]$ posteriors: G-PCA four-parameters case. Same as Figure \ref{fig:mcmc_vs_MC-KL_check_4pars} but for the G-PCA method.
    }
    \label{fig:mcmc_vs_pc_check_4pars}
\end{figure*}

\subsection{PCA + KL}

As described in Paper I, instead of orthogonalising the weights as in \cite{2016PhRvD..93h3525Z}, we perform a principal component analysis (PCA) transformation of our parameter space before applying the KL compression. This is done by diagonalising the Fisher information matrix using the eigenvalue decomposition

\begin{eqnarray}
\mathrm{F}_{\bm{\theta}_{\mathrm{phys.}}} =  \,\bm{\mathrm{P}}\, \;\mathrm{F}_{\bm{\theta}_{\mathrm{PCA}}}\;\bm{\mathrm{P}}^\intercal
\quad
\mathrm{where}
\quad
\bm{\theta}_{\mathrm{PCA}}\,\,=\,\bm{\mathrm{P}}^\intercal\,\bm{\theta}_{\mathrm{phys.}},
\end{eqnarray}

\noindent and $\bm{\mathrm{P}}$ is the linear transformation matrix. After having diagonalised the Fisher matrix we compress the data-vector with respect to this new set of parameters $\bm{\theta}_{\mathrm{PCA}}$. 
The effect of a PCA decomposition is to rotate the parameter space to the axes corresponding to the degeneracies between the original set of parameters. Therefore, taking the outer product of the 1D posteriors of the parameters
$\bm{\theta}_{\mathrm{PCA}}$ in order to get the multidimensional posterior distribution should return a good approximation to the one sampled by the MCMC code.

Since the $\bm{\theta}_{\mathrm{PCA}}$ are uncorrelated, one can randomly sample the 1D posteriors and then rotate the resulting parameter vector using $\bm{\mathrm{P}}$ back into the physical space. Doing this avoids the use of the MCMC sampling altogether.

As shown in Paper I, this works only for those parameter sets which have a sufficiently low degree of degeneracy such that the approximation of Gaussianity for the multidimensional posterior can be assumed to be valid (no or very weak "banana-shaped" contours). Since this is not always the case, as for our choice of parameters, an additional Gaussianisation pre-step is required.

\subsection{Gaussianisation pre-step}
\label{sec:gaussianisation}

In Paper I the PCA + KL method assumed that it was possible to rotate through a linear transformation the physical parameter space into a new one where the new parameters are orthogonal/uncorrelated between each other.
In order to be able to deal with distributions containing non-linear degeneracies (e.g. "banana-shaped" contours),
we add a pre-Gaussianisation transformation of the parameter space using the procedure described in \citet{2016MNRAS.459.1916S}. In their work they introduced an extension of the Box-Cox transformations, which are functions of two parameters $(a,\lambda)$:

\begin{equation}
    \Tilde{\theta}^i = BC_{(a,\lambda)}(\theta^i) = 
        \begin{cases} 
            \lambda^{-1}[(\theta^i+a)^{\lambda}-1] &  (\lambda\neq0) \\ 
            \log(\theta^i+a) &  (\lambda=0)
        \end{cases}
\end{equation}

\noindent where $\Tilde{\theta}^i$ is the transformed $i$-th model parameter while $\theta^i$ is the original $i$-th model parameter.
Their method was labelled Arcsinh-Box-Cox transformation (ABC). For each of the model parameters, a set of three ABC transformation parameters $(a,\lambda,t)$ are computed by the algorithm which are then used in the following way:

\begin{equation}
    \theta^i_{\mathrm{Gauss.}} = ABC(\theta^i_{\mathrm{phys.}}) = 
        \begin{cases} 
            t^{-1}\sinh[t\,BC_{(a,\lambda)}(\theta^i_{\mathrm{phys.}})] &  (t>0) \\ 
            BC_{(a,\lambda)}(\theta^i_{\mathrm{phys.}}) &  (t=0)\\
            t^{-1}\mathrm{arcsinh}[t\,BC_{(a,\lambda)}(\theta^i_{\mathrm{phys.}})] &  (t<0) 
        \end{cases}
\end{equation}

\noindent where $\theta^i_{\mathrm{Gauss.}}$ is the Gaussianised $i$-th model parameter while $\theta^i_{\mathrm{phys.}}$ is the original $i$-th physical model parameter.
We then relabel this compression as G-PCA. In order to obtain the transformation parameters of the Gaussianising transformations it is necessary to run a preliminary MCMC sampling using the full data-vector. What we want to prove is that once the transformation parameters have been obtained for the standard number of triangle bins corresponding to the $\Delta k_6$ binning case, these are valid also for a higher number of triangle bins included in the bispectrum.

\subsection{Analytical covariance matrix: usage}
In the following analysis, we are going to use two different options for the analytical covariance matrices. For the MC-KL method we compress only the bispectrum monopole part of the data-vector. To derive the weights in Eq. \ref{eq:weights} we use the analytical covariance matrix of the bispectrum monopole given by the sum of the Gaussian term in Eq. \ref{eq:cov_b0} and the non-Gaussian one given in Eq. \ref{eq:ng_bb}. 
For the G-PCA method the full data-vector needs to be compressed since the computation of the 1D posteriors of the $\bm{\theta}_{PCA}$ parameters requires each data vector element to be sensitive to the variation of just one $\theta_{PCA}$ parameter, as explained in Paper I. 
Therefore, for the power spectrum monopole/quadrupole we use  Eq. \ref{eq:cov_p02} as our analytical covariance matrix; similarly for the bispectrum monopole we use Eq. \ref{eq:cov_b0} for the covariance matrix (the same as the one we used for the MC-KL case), and finally, we use Eq. \ref{eq:ct_pb} for our cross-covariance matrix.

\section{Recover MCMC-derived posterior distribution}
\label{sec:recover_posterior}

For MCMC sampling we use \textsc{emcee}\footnote{ We use 192 walkers, 1100 burn-in steps and 1700 steps. For the low-resolution MCMC we use half of the previous quantities.}  \citep{2013PASP..125..306F}. All the likelihoods have been corrected as suggested by \cite{2016MNRAS.456L.132S} in order to take into account the bias induced by estimating the inverse of the real covariance matrix from a limited number of mocks. 
In order to check whether our analytical estimate of the covariance matrix is good enough to be used for deriving the weights as explained in Sec. \ref{sec:comp_meth}, we compare to the full MCMC 1D posterior distributions in the left panels of Figures \ref{fig:mcmc_vs_MC-KL_check_4pars} and \ref{fig:mcmc_vs_pc_check_4pars} with results from the MCMC+ MC-KL and G-PCA methods, respectively.

The violin plots include the standard binning case $\Delta k_6$ (116 triangle bins) and the $\Delta k_5$ case (195 triangle bins). For these two cases we compare the MCMC (grey and purple) with the compression results (cyan and orange). From each point we subtract the mean of the model parameters obtained using the MCMC. This makes it easier to check that the shift in the mean of the compression results with respect to the MCMC ones is small when compared to the size of the inner quartiles of the distribution. This concept is also quantified in the bottom half of Table \ref{tab:4pars_consistency}, which shows the shifts in the mean values is relative to the 1D $68\%$ credible intervals. In the top half of Table \ref{tab:4pars_consistency} we report the precise values of both the means and the $68\%$ credible intervals for all model parameters. Additionally, Figure \ref{fig:mcmc_vs_MC-KL_vs_pc_check} in Appendix \ref{sec:validation} shows the comparison between the 2D MCMC posterior distributions and the MC-KL and G-PCA ones for both  $\Delta k_6$ and  $\Delta k_5$ cases. We conclude that even if a small part of the constraining power is lost (see the $\Delta k_6$ columns in Table \ref{tab:4pars_improvement} for details), both compression methods return posterior distributions which well agree with the MCMC distribution for all model parameters under consideration.



\setlength{\tabcolsep}{3pt}


\begin{table*}
  \caption{ Four parameter-case, check consistency. \newline
  \textbf{Upper half:} Mean values of the posterior distributions and $68\%$ credible intervals for the MCMC and the MC-KL / G-PCA compression methods. We report the values for a range of $k$-binnings. From the largest bin $\Delta k_6$, the size used in the BOSS analysis, corresponding to the lowest number of triangle bins (116), to the thinnest binning $\Delta k_2$ corresponding to the highest number of triangle bins (2734). The observed shift in the mean values as a function of the number of triangle bins considered is due to the strong the degeneracy present between the model parameters. As can be seen in Figure \ref{fig:chi_square_4par}, the shift does not have any effect on the goodness of fit. 
  \newline
  \textbf{Lower half:} In the compression columns we report the relative difference between the posterior modes obtained via MCMC and the ones obtained via compression (MC-KL or G-PCA). In the MCMC columns the relative size of the $68\%$ credible intervals obtained via MCMC sampling is shown. Comparing the MCMC columns to the compression ones,  it is that the difference between the mean parameter values obtained via MCMC and the ones obtained via compression (MC-KL or G-PCA) are evidently within the $68\%$ credible intervals given by the MCMC on the full data-vector.}

\makebox[\textwidth][c]{
    \begin{tabular}{ccccccccccc}
\toprule
        & \multicolumn{3}{c}{$\Delta k_6$} & \multicolumn{3}{c}{$\Delta k_5$}& \multicolumn{2}{c}{$\Delta k_4$} & \multicolumn{2}{c}{$\Delta k_2$}  \\
\cmidrule(lr){2-4}\cmidrule(lr){5-7}\cmidrule(lr){8-9}\cmidrule(lr){10-11}
        &  MCMC  & MC-KL  & G-PCA  &  MCMC  & MC-KL  & G-PCA   & MC-KL   & G-PCA  & MC-KL   & G-PCA   \\
\cmidrule(lr){2-4}\cmidrule(lr){5-7}\cmidrule(lr){8-9}\cmidrule(lr){10-11}
$b_1$                 & 2.41 $\pm$ 0.22 & 2.41 $\pm$ 0.23 & 2.49 $\pm$ 0.27 & 2.34 $\pm$ 0.17 & 2.38 $\pm$ \colorbox{red!50!yellow!50!}{0.18} & 2.42 $\pm$ \colorbox{red!50!yellow!50!}{0.17} & 2.27 $\pm$ \colorbox{-red!75!green!50!blue}{0.14} & 2.38 $\pm$ \colorbox{-red!75!green!50!blue}{0.16} &  2.28 $\pm$ \colorbox{-red!75!green}{0.14} & 2.31 $\pm$ \colorbox{red!70}{0.17}\\
$b_2$                 & 1.00 $\pm$ 0.40 & 1.04 $\pm$ 0.42 & 1.08 $\pm$ 0.47 & 0.82 $\pm$ 0.26 & 0.83 $\pm$ \colorbox{red!50!yellow!50!}{0.29} & 0.85 $\pm$ \colorbox{red!50!yellow!50!}{0.26} & 0.79 $\pm$ \colorbox{-red!75!green!50!blue}{0.23} & 0.81 $\pm$ \colorbox{-red!75!green!50!blue}{0.22} & 0.68 $\pm$ \colorbox{-red!75!green}{0.22} & 0.77 $\pm$ \colorbox{red!70}{0.19}\\
$f$                   & 0.69 $\pm$ 0.08 & 0.72 $\pm$ 0.09 & 0.72 $\pm$ 0.09 & 0.67 $\pm$ 0.07 & 0.67 $\pm$ \colorbox{red!50!yellow!50!}{0.07} & 0.70 $\pm$ \colorbox{red!50!yellow!50!}{0.07} & 0.65 $\pm$ \colorbox{-red!75!green!50!blue}{0.06} & 0.68 $\pm$ \colorbox{-red!75!green!50!blue}{0.06} & 0.68 $\pm$ \colorbox{-red!75!green}{0.06} & 0.67 $\pm$ \colorbox{red!70}{0.06}\\
$\sigma_8$            & 0.50 $\pm$ 0.04 & 0.48 $\pm$ 0.05 & 0.48 $\pm$ 0.05 & 0.51 $\pm$ 0.04 & 0.50 $\pm$ \colorbox{red!50!yellow!50!}{0.04} & 0.49 $\pm$ \colorbox{red!50!yellow!50!}{0.03} & 0.53 $\pm$ \colorbox{-red!75!green!50!blue}{0.03} & 0.51 $\pm$ \colorbox{-red!75!green!50!blue}{0.03} & 0.52 $\pm$ \colorbox{-red!75!green}{0.03} & 0.51 $\pm$ \colorbox{red!70}{0.03}\\
\cmidrule(lr){2-4}\cmidrule(lr){5-7}\cmidrule(lr){8-9}\cmidrule(lr){10-11}
& $\dfrac{\Delta\theta^{\mathrm{mc}}_{\Delta k_6}}{\theta^{\mathrm{mc}}_{\Delta k_6}}\;\left[\%\right]$ 
& \multicolumn{2}{c}{$\dfrac{\theta^{\mathrm{comp.}} - \theta^{\mathrm{mc}}_{\Delta k_6}}{\theta^{\mathrm{mc}}_{\Delta k_6}}\;\left[\%\right]$}
& $\dfrac{\Delta\theta^{\mathrm{mc}}_{\Delta k_5}}{\theta^{\mathrm{mc}}_{\Delta k_5}}\;\left[\%\right]$ 
& \multicolumn{2}{c}{$\dfrac{\theta^{\mathrm{comp.}} - \theta^{\mathrm{mc}}_{\Delta k_5}}{\theta^{\mathrm{mc}}_{\Delta k_5}}\;\left[\%\right]$}
& \multicolumn{2}{c}{$\dfrac{\theta^{\mathrm{comp.}} - \theta^{\mathrm{mc}}_{\Delta k_5}}{\theta^{\mathrm{mc}}_{\Delta k_5}}\;\left[\%\right]$}
& \multicolumn{2}{c}{$\dfrac{\theta^{\mathrm{comp.}} - \theta^{\mathrm{mc}}_{\Delta k_5}}{\theta^{\mathrm{mc}}_{\Delta k_5}}\;\left[\%\right]$}  \\
\cmidrule(lr){2-2}\cmidrule(lr){3-4}\cmidrule(lr){5-5}\cmidrule(lr){6-7}\cmidrule(lr){8-9}\cmidrule(lr){10-11}
$ b_1 $      & 9.2  & -0.3 & 3.3  & 7.3  & 1.9   & 3.5  & -2.7  & 1.9  &  -2.7  & -1.1 \\
$ b_2 $      & 40.3 &  3.5 & 7.5  & 32.2 & 1.9   & 4.4  & -3.6  & -1.2 &  -16.5 & -5.7 \\
$ f   $      & 12.1 &  4.4 & 4.4  & 10.1 & -1.3  & 3.8  & -3.3  & 0.2  &   0.5  & -1.1 \\
$ \sigma_8$  & 8.5  & -5.1 & -5.5 & 7.3  & -1.1  & -3.6 &  4    & -0.3 &  2.2   & -1.2 \\
\bottomrule
\end{tabular}%
}
\label{tab:4pars_consistency}%
\end{table*}%


\begin{table*}
  \caption{ Four-parameter case, constraints improvement. Below are shown the relative variations in percentage of the size of the $68\%$ credible intervals as a function of the $k$-binning considered (number of triangle bins used for the bispectrum monopole). In orange and green are highlighted respectively the improvements achieved via compression for the $\Delta k_5$ and at the saturation level (404 triangle bins - $\Delta k_4$) of the bispectrum monopole constraining power case for the considered set of parameters (e.g. left panel of Figure \ref{fig:oned_improv_and_onesigma_shift}). Finally in blue and red are highlighted the improvements obtained via compression for the highest number of triangle bins considered (2734 triangle bins - $\Delta k_2$ binning) for MC-KL and G-PCA respectively.
  }
\makebox[\textwidth][c]{
    \begin{tabular}{ccccccccccc}
\toprule
        & \multicolumn{3}{c}{$\Delta k_6$} & \multicolumn{3}{c}{$\Delta k_5$}& \multicolumn{2}{c}{$\Delta k_4$} &  \multicolumn{2}{c}{$\Delta k_2$}  \\
\cmidrule(lr){2-4}\cmidrule(lr){5-7}\cmidrule(lr){8-9}\cmidrule(lr){10-11}
& $\Delta\theta^{\mathrm{mc}}_{\Delta k_6}$ 
& \multicolumn{2}{c}{$\dfrac{\Delta\theta^{\mathrm{comp.}} - \Delta\theta^{\mathrm{mc}}_{\Delta k_6}}{\Delta\theta^{\mathrm{mc}}_{\Delta k_6}}\;\left[\%\right]$}
& $\Delta\theta^{\mathrm{mc}}_{\Delta k_5}$
& \multicolumn{2}{c}{$\dfrac{\Delta\theta^{\mathrm{comp.}} - \Delta\theta^{\mathrm{mc}}_{\Delta k_6}}{\Delta\theta^{\mathrm{mc}}_{\Delta k_6}}\;\left[\%\right]$}
& \multicolumn{2}{c}{$\dfrac{\Delta\theta^{\mathrm{comp.}} - \Delta\theta^{\mathrm{mc}}_{\Delta k_6}}{\Delta\theta^{\mathrm{mc}}_{\Delta k_6}}\;\left[\%\right]$}
& \multicolumn{2}{c}{$\dfrac{\Delta\theta^{\mathrm{comp.}} - \Delta\theta^{\mathrm{mc}}_{\Delta k_6}}{\Delta\theta^{\mathrm{mc}}_{\Delta k_6}}\;\left[\%\right]$}  \\
\cmidrule(lr){2-2}\cmidrule(lr){3-4}\cmidrule(lr){5-5}\cmidrule(lr){6-7}\cmidrule(lr){8-9}\cmidrule(lr){10-11}
        &  MCMC  & MC-KL  & G-PCA  &  MCMC  & MC-KL  & G-PCA   &   MC-KL   & G-PCA  &  MC-KL   & G-PCA   \\
\cmidrule(lr){2-2}\cmidrule(lr){3-4}\cmidrule(lr){5-5}\cmidrule(lr){6-7}\cmidrule(lr){8-9}\cmidrule(lr){10-11}
$\Delta b_1 $     & 0.22 & 4.4 & 18.8 & 0.17 & \colorbox{red!50!yellow!50!}{-21.3} & \colorbox{red!50!yellow!50!}{-22.0}   & \colorbox{-red!75!green!50!blue}{-35.3} & \colorbox{-red!75!green!50!blue}{-30.0}   &  \colorbox{-red!75!green}{-35.3} & \colorbox{red!70}{-24.8} \\
$\Delta b_2 $     & 0.40 & 2.9 & 16.2 & 0.26 & \colorbox{red!50!yellow!50!}{-28.9} & \colorbox{red!50!yellow!50!}{-35.0}   & \colorbox{-red!75!green!50!blue}{-42.6} & \colorbox{-red!75!green!50!blue}{-46.0}   & \colorbox{-red!75!green}{-45.3} & \colorbox{red!70}{-52.8} \\
$\Delta f   $     & 0.08 & 3.7 & 7.0    & 0.07 & \colorbox{red!50!yellow!50!}{-16.5} & \colorbox{red!50!yellow!50!}{-18.5} & \colorbox{-red!75!green!50!blue}{-24.7} & \colorbox{-red!75!green!50!blue}{-25.1} & \colorbox{-red!75!green}{-22.6} & \colorbox{red!70}{-26.4} \\
$\Delta \sigma_8$ & 0.04 & 6.5 & 10.0   & 0.04 & \colorbox{red!50!yellow!50!}{-11.3} & \colorbox{red!50!yellow!50!}{-18.7} & \colorbox{-red!75!green!50!blue}{-22.3} & \colorbox{-red!75!green!50!blue}{-24.5} &  \colorbox{-red!75!green}{-22.6} & \colorbox{red!70}{-21.0}   \\
\bottomrule
\end{tabular}%
}
\label{tab:4pars_improvement}%
\end{table*}%


\section{Information content and number of triangle bins}
\label{sec:info_tr_num}
The right panels of  Figures \ref{fig:mcmc_vs_MC-KL_check_4pars} and \ref{fig:mcmc_vs_pc_check_4pars} show how using a $\sim23$ times larger number of triangle bins tightens the posterior contours of the four model parameters considered and reduces the degeneracies between them. At the same time, the maxima of the 2D posterior distributions converge to the same values for each compression method as the number of triangle bins is increased.

Note that the shift in the posterior distribution between binning cases is not an artifact of the compression: it is also present when when we fit using the standard MCMC method. 
This can be seen when comparing the location and shape of the 2D contour regions in Figures \ref{fig:mcmc_vs_MC-KL_vs_pc_check} and \ref{fig:mcmc_vs_MC-KL_vs_pc_check2} in Appendix \ref{sec:validation} for the $\Delta k_6$ and $\Delta k_5$ binning cases. Quantitatively it can be observed by comparing means and standard deviations in Table \ref{tab:4pars_consistency}. Thus, both compression algorithms 
reproduce posterior distributions very similar to the ones derived via MCMC sampling
for the relevant binning cases  $\Delta k_6$ and $\Delta k_5$. 
The observed shift between binning cases is due to the strong degeneracy between the model parameters. In particular the shift happens along the degeneration direction of $b_1$, $b_2$ and $f$ with $\sigma_8$. It may have a statistical origin. Further checks on this effect may be performed using the galaxy mocks, for example by fitting several different realizations for both the $\Delta k_6$ and $\Delta k_5$ binning cases using the G-PCA method (which would be much faster than doing parameter estimation via MCMC or MC-KL). We reserve to do these tests in future work.
Additionally, the practically identical (compared to the errorbars amplitude) residuals plots for the different models in Figure \ref{fig:chi_square_4par} show that the shifts in the best-fit parameters as a function of the number of triangle bins used is an effect of the strong degeneracy present in the parameter space. Even if employing more triangle bins partially lifts this, the degree of how well the models for the different number of triangle bins fit the data does not change.

\begin{figure*}%
    \centering
    \subfloat[information content vs. triangle number]{{\includegraphics[width=0.5\textwidth]
    {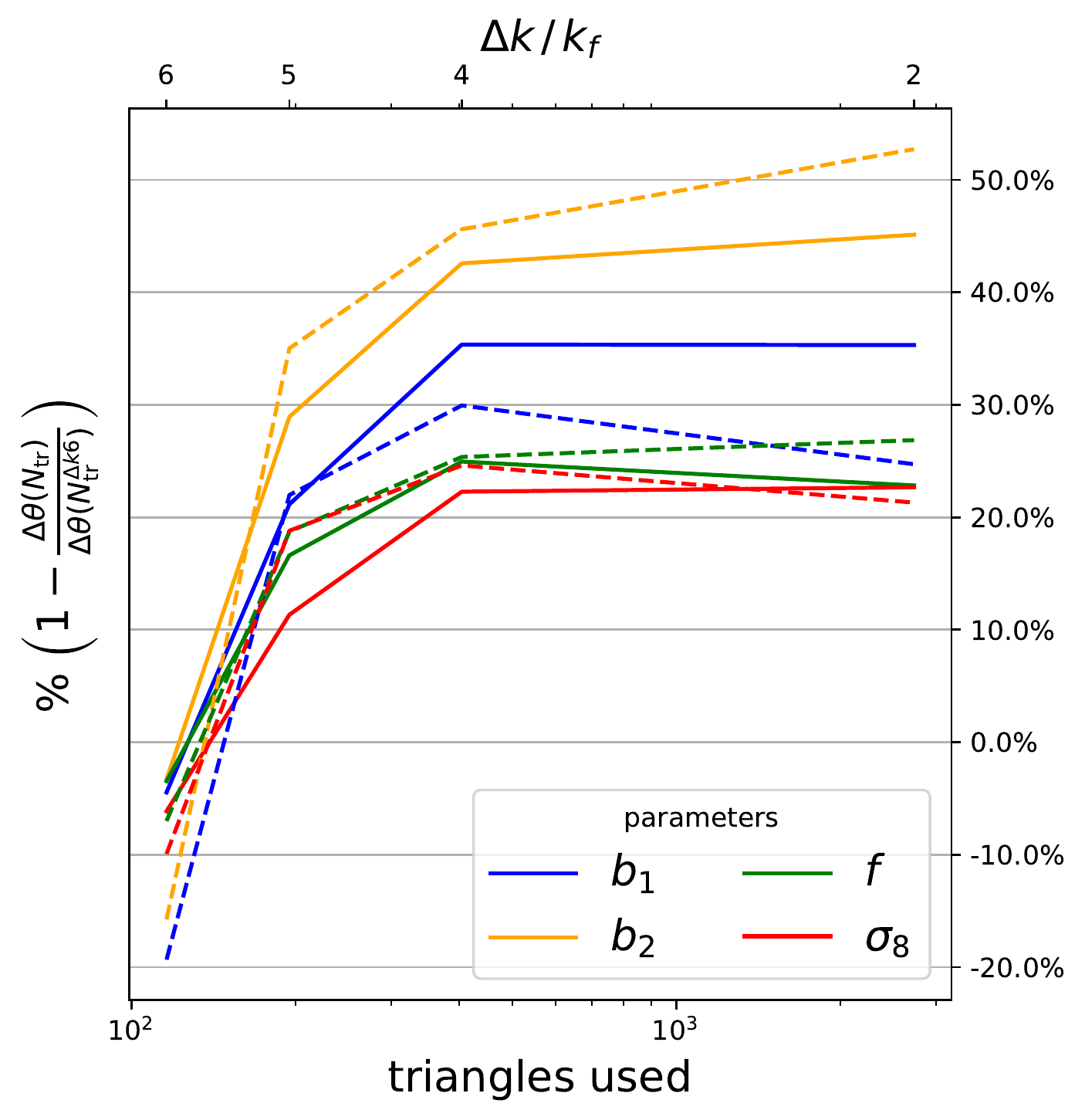}
    }}%
    \subfloat[shifted fiducial parameter set]{{\includegraphics[width=0.5\textwidth,height=0.5\textwidth]
    {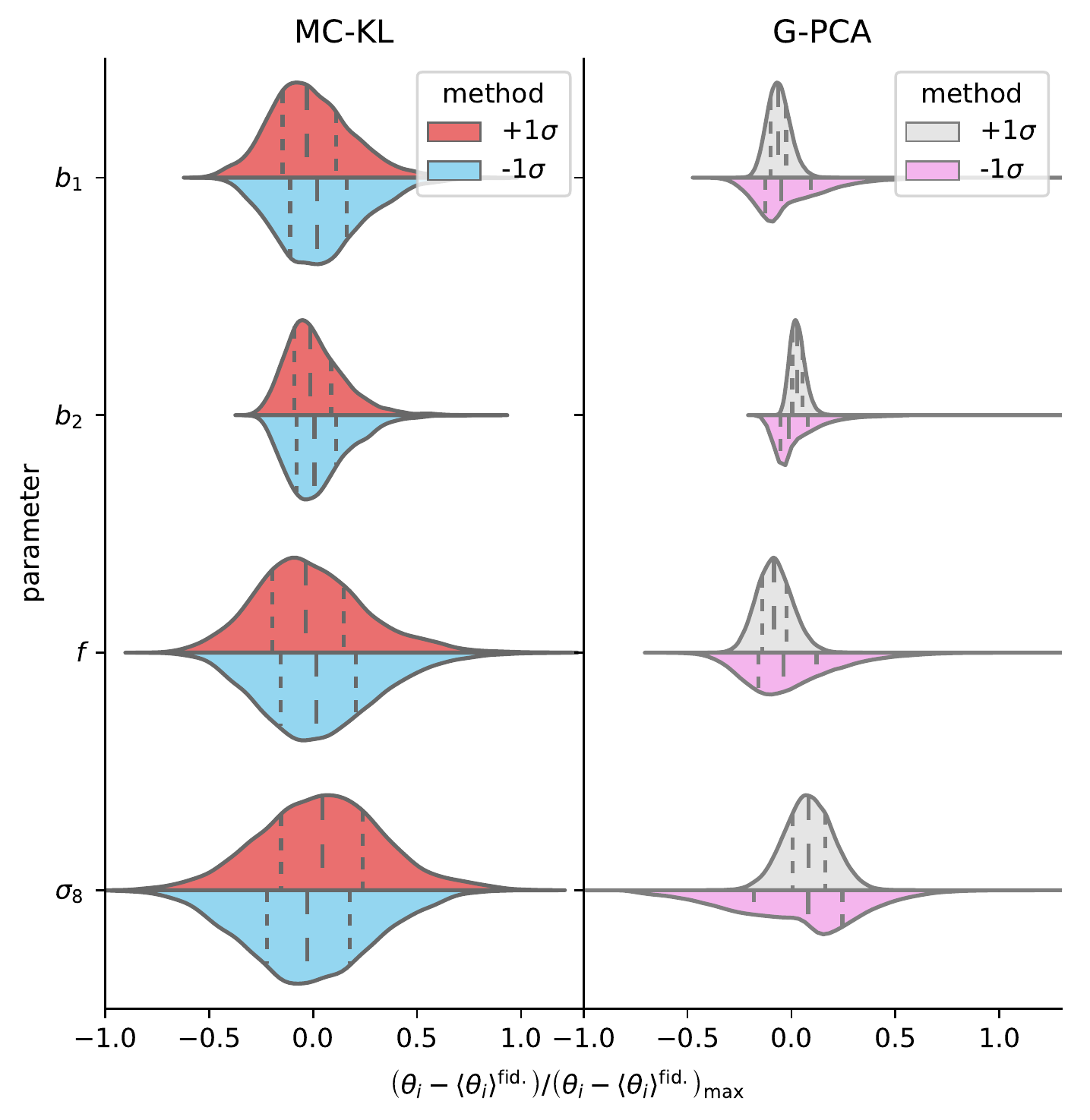}
    }}%
    \caption{
    \textbf{a)} the 1D $68\%$ credible intervals as a function of the number of triangle bins used in the bispectrum monopole data-vector. Continuous lines represent the MC-KL results while the dashed ones are given by the G-PCA compression method. 
    \newline
   \textbf{b)} the compression results for the MC-KL and G-PCA cases when the fiducial parameter set used to compute the analytical covariance matrix and the derivatives of the mean are shifted by $\pm 1 \sigma$ credible intervals. The violin plots show, for the test case of the $\Delta k_6$-binning, the comparison between the 1D posterior distributions for all parameters, using  shifts by $+1 \sigma$ (red/grey) and $-1\sigma$ (blue/pink)  for the MC-KL / G-PCA methods. The vertical lines represent the $25\%$, $50\%$ and $75\%$ quartiles. All distributions are mean-subtracted using the fiducial parameter set for the compression, and they have been normalised by the maximum difference between the parameter value of each sample and the mean of the distribution. Even if the 1D distributions are not Gaussian, the effect of compressing with respect a shifted cosmology is qualitatively negligible for the MC-KL method while it affects the G-PCA performance more. Nevertheless, the modifications to the fiducial parameter sets are substantial ($\sim 10 -40 \%$ varations) given the broad posteriors due to the strong degeneracy in the parameter set.}
    \label{fig:oned_improv_and_onesigma_shift}
\end{figure*}


The main result of this paper is that the variance of the parameters is substantially reduced when the number of triangle bins used is increased up to $\sim23$ times the original number.
In terms of percentages of the original 1D $68\%$ credible intervals obtained running an MCMC on the full data-vector for the parameters $\left(b_1,b_2,f,\sigma_8\right)$ in the BOSS $\Delta k_6$ case, the $\Delta k_2$  MC-KL and G-PCA analyses obtain tighter constraints by $\left(-35\%,-45.3\%,-22.6\%,-22.6\%\right)$ and $\left(-24.8\%,-52.8\%,-26.4\%,-21\%\right)$, respectively. These optimal constraints as obtained by the compression methods are also shown in summary in Figure \ref{fig:conc_best_const}.
The gain in parameter constraints is due to the fact that when we increase the number of triangle bins, by decreasing the $k$-bins size, the information is less “washed out” than when using larger $k$-bins.

For future surveys the compression can be then used to maximise the constraining power of the main analysis and also to find out the minimum number of triangle bins for a given $k$-range needed to fully capture the non-Gaussian information contained in 3pt statistics like the bispectrum. The later will indicate how many mock catalogues/simulations are required in order to accurately estimate the covariance matrix. In our analysis the saturation seems to be reached already for the $\Delta k_4$ binning case (404 triangle bins).

For what concerns $\Delta k_2$, the smallest $k$-bin size considered (2734 triangle bins), Tables \ref{tab:4pars_consistency} and \ref{tab:4pars_improvement} show that the $\Delta k_2$ posterior distribution is very similar to the $\Delta k_4$ case.

The trend in the information content in terms of the 1D $68\%$ credible intervals as a function of the triangle number used is shown in the left panel of Figure \ref{fig:oned_improv_and_onesigma_shift}, and the improvement quantified in Table \ref{tab:4pars_improvement}. From Figure \ref{fig:oned_improv_and_onesigma_shift} it appears that the parameters constraints improvement as a function of the number of triangle bins reaches the saturation already for the $\Delta k_4$ case.
For the chosen $k$-range, the additional triangle bins (and bispectra) included in the $\Delta k_2$ with respect to the $\Delta k_4$ one do not substantially add new features to the bispectrum data-vector, therefore the constraining power results weakly improved.


\begin{figure*}
\centering
\includegraphics[width=\textwidth]{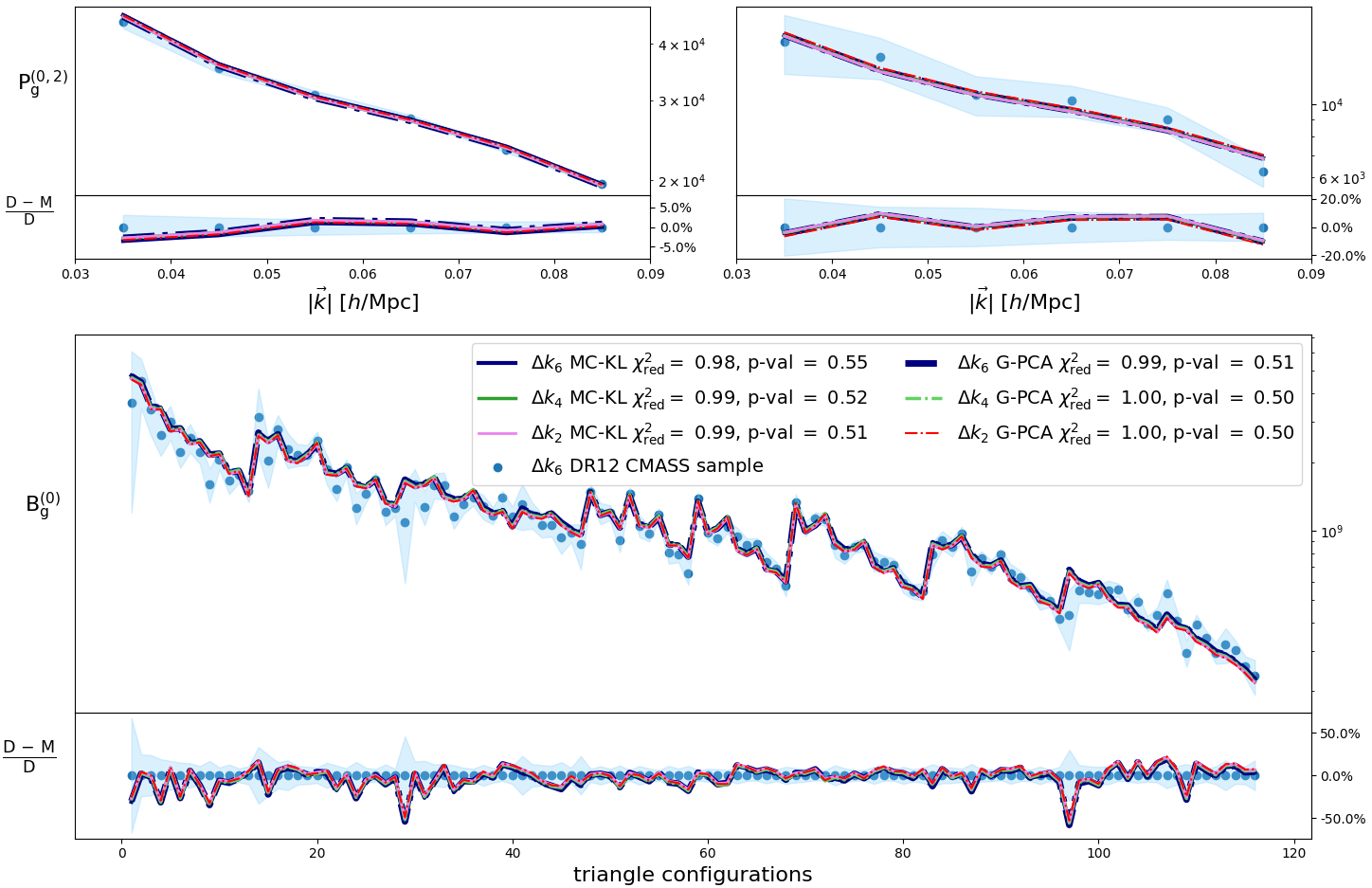}
\caption{Reduced $\chi^2$ and $p$-values for the best-fit models obtained using the MCMC, MC-KL and G-PCA compression methods.
The $k$-binnings shown are respectively the standard $\Delta k_6$ (navy), an intermediate size $\Delta k_4$ (green) and the smallest one $\Delta k_2$ (pink for MC-KL and red for G-PCA) corresponding to the highest number of triangle used in the bispectrum monopole. The two upper panels are for the power spectrum monopole (left) and quadrupole (right) while the bottom panel refers to the bispectrum monopole. The lower part of each panel shows the relative difference between the data measurements and the different models. Even if for example $b_1$ and $\sigma_8$ values are shifted between the cases of $\Delta k_6$ and $\Delta k_2$, the strong degeneracy has the result of making the two models practically identical.}
\label{fig:chi_square_4par}
\end{figure*}


\section{consistency check}
\label{sec:consistency_check}
In order to test the validity of our analysis, we compute the reduced $\chi^2$ and corresponding $p$-value for each set of parameters obtained using either the MCMC sampling or the compression methods. For all parameter vectors (compressed and uncompressed) this has been done using the data-vector corresponding to the standard $\Delta k_6$ binning. The results can be seen in Figure \ref{fig:chi_square_4par}. This test proves that the shift observed in the parameters as the number of triangle bins is increased is simply due to the strong degeneracy present between $b_1$, $b_2$, $f$ and $\sigma_8$. Indeed both the reduced $\chi^2$ and $p$-values show that all these models fit the data very well. In Figure \ref{fig:chi_square_4par} we did not show the lines and statistics for the 
$\Delta k_5$ cases just for the sake of clarity and because the results are equivalent to those of the other binnings. From the same figure it can also be noticed that the tightest errorbars are those from the power spectrum case. 

To demonstrate the flexibility of the compression methods we check their performance when the fiducial parameter set is shifted by $\pm1\sigma$ credible intervals in the $\Delta k_6$ case. The effect of this is shown in the right panel of Figure \ref{fig:oned_improv_and_onesigma_shift}. For this plot, we centre each 1D distribution by subtracting the mean obtained by running the compression pipelines using the fiducial parameters values. In this way it is possible to observe by how much the posterior distributions derived via MC-KL or G-PCA shift as a function of the chosen fiducial parameter set. In Appendix \ref{sec:validation} the precise numbers are reported in Table \ref{tab:4pars_shifted_cosmology}.

MC-KL appears to be more stable than the G-PCA when the fiducial parameter set is shifted. The explanation of this could be the fact that G-PCA involves several transformations of the parameter space, including a diagonalisation of the Fisher information matrix which is computed from the analytical model of the covariance matrix. 

Nevertheless, it should be noted that we are testing the performances of the compression in a regime of strong degeneracy of the parameter space and therefore shifting the fiducial parameter set by $\pm1\sigma$ credible intervals actually means increasing/reducing the individual values by $\sim 10 -40 \%$ (second panel Table \ref{tab:4pars_consistency}). Therefore, running a preliminary low-resolution MCMC sampling on the full data-vector (which can be shorter than the one that will be later compressed, as we have done in our analysis) is an efficient solution to determining a reasonable fiducial model for deriving the compression.

\subsection{Comparison with BOSS DR12 bias constraints}
BOSS galaxy sample results from the bispectrum are reported by \citet{2017MNRAS.465.1757G}  [in Table 3 at p. 18] from the same CMASS sample data set, at the same redshift, for the following parameter combinations: $b_1\sigma_8 = 1.2479\pm0.0072$, $\,b_2\sigma_8= 0.641\pm 0.066$ and $f\sigma_8=0.432\pm0.018$\footnote{we compare our results with the BOSS analysis standard deviation values obtained considering only the statistical contributions and not the systematics ones.}. If we recast our results obtained using the MCMC for the $\Delta k_6$ case in terms of the same parameter combinations these are: $b_1\sigma_8 = 1.203\pm0.008$, $b_2\sigma_8 = 0.557\pm0.140$ and $f\sigma_8 = 0.339\pm0.019$.

In the BOSS analysis a larger range of scales has been considered. In particular, BOSS analysis goes up to $k\sim 0.2h/\mathrm{Mpc}$ for both power spectrum monopole/quadrupole and bispectrum monopole while we stop at $k\sim 0.09 h/\mathrm{Mpc}$ and $k\sim 0.12 h/\mathrm{Mpc}$, respectively. This could explain the larger value we obtained for $b_2\sigma_8$. A more complex model for the power spectrum was used in the BOSS analysis, including loop corrections beyond the tree level approximation. Moreover the BOSS analysis also modelled the effect of the survey window function for both power spectrum and bispectrum. 

As we saw from Figure \ref{fig:chi_square_4par}, the power spectrum monopole is the most constraining part of the full data-vector, having errorbars of less than $5\%$. 
Therefore it is possible that our simple tree-level approximation for the power spectrum, besides limiting the $k$-range analysed, could be the cause of the discrepancy between the BOSS results with respect to the relative lower values obtained for the combined parameters $b_1\sigma_8$, $b_2\sigma_8$ and $f\sigma_8$ in this work.

Moreover, in the BOSS analysis the FoG parameters $\sigma_{\mathrm{FoG}}^{\mathrm{B}} $ and $\sigma_{\mathrm{FoG}}^{\mathrm{P}}$ were left free to vary in order to better model the non-linear regime and were detected with high significance ($\sigma_{\mathrm{FoG}}^{\mathrm{B}} = 7.54\pm0.70 $ and $\sigma_{\mathrm{FoG}}^{\mathrm{P}}=3.50\pm0.14$). The BOSS model also included a noise-amplitude parameter $A_{\mathrm{noise}}$ which modelled divergence from Poissonian shot noise. In our model we had included $A_{\mathrm{noise}}$ initially, however we set it to zero after having checked that, if let free to vary, its posterior distribution was compatible with zero for the $k$-range considered. These differences in the modelling and scales considered could explain the discrepancy in the best-fitting parameters.

In Appendix \ref{sec:validation} and in particular Figure \ref{fig:fixed_s8_chi2}, we show test the limitation to the data-vector constraining power, implied by our choice of $k$-range, by running an MCMC sampling for two parameter sets: $\left(b_1,b_2,f,\sigma_8\right)$ and $\left(b_1,b_2,f\right)$. In the second case, by fixing $\sigma_8$ to its fiducial value, we recover maximum likelihood values for  $\left(b_1,b_2,f\right)$
very different from the ones corresponding to the four parameter case reported in Table \ref{tab:4pars_consistency} $\left(1.98\pm0.01,0.39\pm0.06,0.53\pm0.03\right)$. However this discrepancy is not reflected in the reduced $\chi^2$ values for the two different sets of best-fit parameters: for the $\left(b_1,b_2,f,\sigma_8\right)$ case $\chi^2_{\mathrm{red}}=0.98$ while for the set $\left(b_1,b_2,f\right)$ $\chi^2_{\mathrm{red}}=1.05$. The fact that both best-fit parameter sets very well fit the data implies that the constraining power of the data-vector on the data in the four parameter case is not sufficient to lift the degeneracies present in the parameter space. Therefore in order to lift the degeneracies and to avoid a shift in the inferred parameters when $\sigma_8$ is also fitted, a more accurate model for the power spectrum monopole and quadrupole including loop correction is needed.

\subsection{Difference in time and computer resources needed}

There is no significant difference between MCMC and MC-KL in terms of time taken for the pipeline to run or computing resources needed. For the parameter set $\left(b_1,b_2,f,\sigma_8\right)$ the running time varied between 20 minutes for 116 triangle bins to $\sim 10$ hours for 2734 triangle bins on 14 2.2 GHz Intel i7 cores. G-PCA proved to be faster when many triangle bins are used. Considering $\sim 30$ minutes for the preliminary MCMC with 116 triangle bins and $\sim$ 2 hours for the Gaussianisation part, it took between $\sim$5 minutes (116 triangle bins) and $\sim 30$ minutes (2734 triangle bins) using only one 2.2 GHz Intel i7 core for the compression plus posterior evaluation to run. Therefore, by running once the preliminary MCMC and Gaussianisation algorithm, we were able to run the PCA part for all the binning cases considered in less than in total $\sim$ 3 hours wall-clock time.

We used CAMB \citep{2000ApJ...538..473L} to compute the linear matter power spectrum.
The time difference between MCMC/MC-KL and G-PCA would have been much more significant in the case of a parameter set for which the linear matter power spectrum needs to be recomputed for every model realisation.


\begin{figure*}
\centering
\includegraphics[width=\textwidth]
{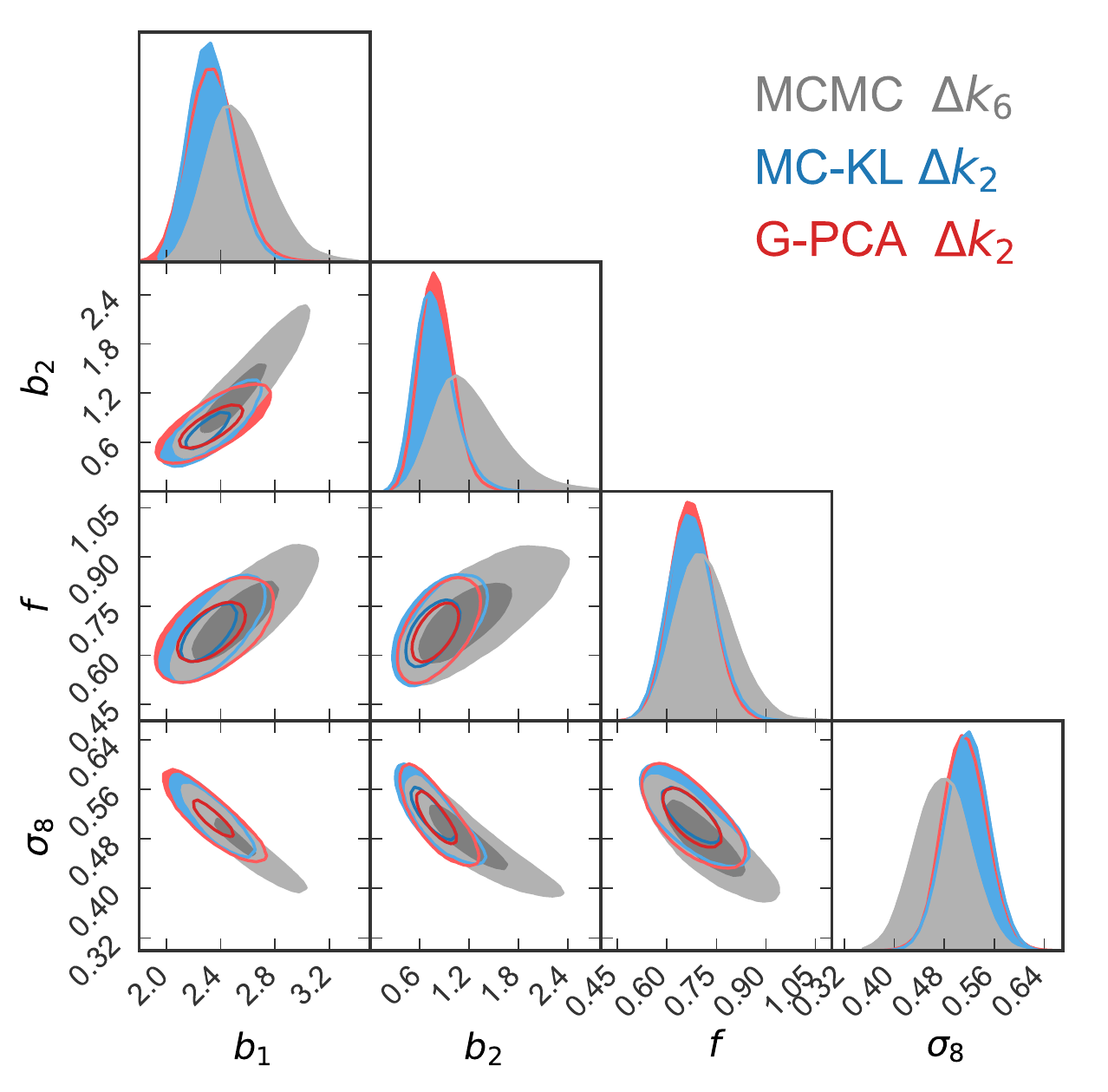}
\caption{MCMC vs. MC-KL vs. G-PCA. 2-D  $68\%$ and $95\%$ credible contours are shown respectively for the $\Delta k_6$ MCMC (grey), $\Delta k_2$ MC-KL (blue) and $\Delta k_2$ G-PCA (red) cases. It is possible to observe the substantial improvement in parameter constraints through applying either compression method to a data-vector containing approximately $\sim23$ times more triangle bins than the one used for the MCMC sampling case. The agreement between the MC-KL and G-PCA posterior distributions is remarkable. Using more triangle bins helps with lifting the strong degeneracy between the model parameters, as can be seen from the shrinkage of the 2-D contours along the degeneracy directions.}
\label{fig:conc_best_const}
\end{figure*}


\section{Conclusions}
\label{sec:conclusions}
In this paper we have shown the results of applying both compression methods for the galaxy redshift-space bispectrum, presented in Paper I, to the measurements from the SDSS-III BOSS DR12 CMASS sample \citep{2017MNRAS.465.1757G}. We considered as original data-vector the combination of the power spectrum monopole and quadrupole with the bispectrum monopole, which are obtained by averaging over the angles describing the orientation with respect to the line of sight. The first method called MC-KL consists of running an MCMC sampling on the compressed data-vector obtained by taking the scalar product between the original data-vector and a set of weights derived as first shown by \citet{1997ApJ...480...22T}. The second method, which we denoted as G-PCA, is the modification of the PCA + KL method presented in Paper I obtained by adding a Gaussianisation transformation of the parameter set \citep{2016MNRAS.459.1916S} before rotating it using a principal component analysis transformation (PCA) followed by the KL compression.
By transforming the physical parameter space into an orthogonal one it is possible to just randomly sample 1D posterior distributions, avoid altogether the need of running a MCMC routine.

In order to derive the posterior distributions for the set of parameter considered, the galaxy bias parameters $b_1$ and  $b_2$, the growth rate $f$ and the normalisation of the dark matter perturbations amplitude $\sigma_8$, we numerically estimated the covariance matrix using 1400 and 700 galaxy mocks catalogues for the full data-vector and compressed data-vector cases, respectively.

The following points represent the main conclusions of our analysis:

\begin{itemize}
    \item In order to obtain the weights for the compression methods we derived an analytic approximation of the leading terms of the covariance matrix relative to the considered data-vector. The final expressions of these computations are reported in Sec. \ref{sec:datav_cov} while the full derivations are shown in Appendix \ref{sec:est_def}.
    
    \item In Sec. \ref{sec:recover_posterior} we have shown that both compression methods recover the posterior distributions obtained via MCMC using the full data-vector with little loss of information ($\sim4\%$ and $\sim13\%$ larger 68$\%$ credible intervals than the MCMC ones in average for MC-KL and G-PCA, respectively). More importantly, even if slightly broader, the posterior distributions recovered through compression have the same shape and modes as the MCMC counterparts. 
    
    \item Adding a pre-Gaussianisation step removes the PCA + KL limitation linked to a strongly degenerate parameter space described in Paper I. It is however necessary to run a preliminary MCMC in order to derive the Gaussianisation transformation parameters. Nevertheless, once these parameters have been derived for a number of triangle bins case for which it is possible to run an MCMC on the full data-vector, they can then be used to compress a data-vector with an arbitrary number of triangle bins.
    The decrease in the compression perfomances shown in Figure \ref{fig:oned_improv_and_onesigma_shift} due to a far from optimal choice of fiducial model parameters is also solved by re-running the compression using as fiducial model the parameters inferred in the first run.
    
    \item In Sec. \ref{sec:info_tr_num} we show the main result of this work, namely the substantial improvement in parameter constraints obtained by compressing a much larger number of triangle bins with respect to standard MCMC data-vector. For the uncompressed data-vector the number of triangle bins is limited by the number of mock catalogues available to estimate the covariance matrix. For both compression methods and for any number of triangle configuration considered, the dimension of the compressed data-vector is always equal to the number of model parameters constrained. 
    
    For the highest number of triangle bins considered, this leads to an improvement in terms of the $68\%$ 1D credible intervals by $\left(-35\%,-45\%,-23\%,-23\%\right)$ and $\left(-25\%,-53\%,-26\%,-21\%\right)$ for the MC-KL and G-PCA methods, respectively.
    
    \item By way of summary, in Figure \ref{fig:conc_best_const} we show the results for both MC-KL and G-PCA methods using 2734 triangle bins and for the MCMC on the uncompressed data-vector containing 116 triangle bins.
    The two compression methods agree well and produce substantially tighter and less degenerate constraints.
    Furthermore the G-PCA approach allowed for a computational speed up, requiring only approximately a third of the time taken by the MCMC and MC-KL methods, including also the low-resolution MCMC necessary for the Gaussianisation transformation. Considering only the PCA part, the speed up factor rises to $\sim 20-100$ times depending on the parameter set considered.
    
    \item Finally we would like to point out that the compressing methods used in this work represents a straightforward approach to include higher order statistics like the trispectrum or the tetraspectrum in the analysis of current and future data sets. This is due to the fact that the number of elements of the data-vector, after the maximal compression, corresponds exactly to the number of model parameters. 
    Both MC-KL and G-PCA have the potential to fully exploit the constraining power of higher order statistics applied to data-sets from future surveys like DESI, EUCLID and PFS. 
    
\end{itemize}

\section*{Acknowledgements} 
D.G. is grateful to L. Whiteway for the useful discussions and to E. Edmondson for the technical help in using the UCL computer cluster.
D.G. is supported by the Perren and the IMPACT studentships.
HGM is supported by Labex ILP (reference ANR-10-LABX-63) part of the Idex SUPER, and received financial state aid managed by the Agence Nationale de la Recherche, as part of the programme Investissements d'avenir under the reference ANR-11-IDEX-0004-02.e
M.M. acknowledges funding from STFC Consolidated Grants RG84196 and
RG70655 LEAG/506 and has received funding from the European Union’s Horizon 2020 research and innovation programme under Marie Skłodowska-Curie grant agreement No 6655919.
O.L. acknowledges support from a European Research Council Advanced Grant FP7/291329.
C \citep{Kernighan:1988:CPL:576122} and \textsc{python} 2.7 \citep{Rossum:1995:PRM:869369} have been used together with many packages like I\textsc{python} \citep{Perez:2007:ISI:1251563.1251831}, Numpy \citep{DBLP:journals/corr/abs-1102-1523}, Scipy \citep{jones} and Matplotlib \citep{Hunter:2007:MGE:1251563.1251845}. The corner plots have been realised using \textsc{pygtc} developed by \citet{Bocquet2016} while the violin plots have been created using \textsc{seaborn} by \citet{michael_waskom_2016_45133} .




\bibliographystyle{mnras}


\bibliography{ads_reference.bib}


\end{multicols*}

\appendix

\section{Estimators and covariance terms}
\label{sec:est_def}

\subsection{Power spectrum monopole/quadrupole and bispectrum monopole estimators}
\label{sec:estimators}

The computations for the power spectrum and bispectrum multipoles below reported are original of this work. Expressions for the matter power spectrum and bispectrum were derived also by \citet{1998ApJ...496..586S,2006PhRvD..74b3522S}, however in this work we proceed similarly to what done by \citet{2013MNRAS.429..344K}.

The analytical model for the redshift-space galaxy power spectrum monopole and quadrupole is given by equation \ref{pk_02_esp}.

\noindent It is therefore natural to define the estimator as:

\begin{eqnarray}
\label{pk_02_est}
\hat{\mathrm{P}}^{(\ell)}_{\mathrm{g}}\left(k\right) = 
\left(\dfrac{2\ell+1}{2}\right)\dfrac{1}{(2\pi)^3N_{\mathrm{p}}(k)}\int_{V_{\bm{p}}}\int_{V_{\bm{q}}}\,d^3\bm{p}d^3\bm{q}
\, L_{\ell}\left(\mu\right)
\delta_\mathrm{D}\left(\bm{q}+\bm{p}\right)
\delta_{g}^{\mathrm{s}}\left(\bm{q}\right)\delta_{g}^{\mathrm{s}}\left(\bm{p}\right),
\end{eqnarray}

\noindent where $V_{\bm{p},\bm{q}}$ are the spherical shell volumes characterised by $k-\Delta k/2 \leq q,p \leq k+\Delta k/2 $. $\mu$ is the cosine of the angle with respect to the line of sight of the $\bm{q}$ wave vector and $L_{\ell}\left(\mu\right)$ is the Legendre polynomial of order $\ell$. $\delta_D$ is the 3-D Dirac delta. $N_{\mathrm{p}}$ is the number of grid point pairs in the integration volume in Fourier space and can be computed as:

\begin{eqnarray}
\label{eq:n_pairs}
N_{\mathrm{p}}(k) = \dfrac{V_k}{k_f^3} = k_f^{-3}\,\int_{V_{\bm{p}}}\int_{V_{\bm{q}}}\,d^3\bm{p}d^3\bm{q}\delta_\mathrm{D}\,\left(\bm{q}+\bm{p}\right) \simeq \dfrac{ 4\pi k^2\Delta k}{k_f^3},
\end{eqnarray}

\noindent where $V_k \simeq 4\pi k^2\Delta k $ is the spherical integration shell defined by $k-\Delta k/2 \leq q,p \leq k+\Delta k/2 $ as defined in \citet{1998ApJ...496..586S}. $k_f$ is the fundamental frequency defined in terms of the survey volume $V_{\mathrm{e}}$ as $k_f^3 = \dfrac{(2\pi)^3}{V_{\mathrm{e}}}$. We check that the estimator defined in Eq. \ref{pk_02_est} is unbiased:

\begin{align}
    \langle \hat{\mathrm{P}}^{(\ell)}_{\mathrm{g}}\left(k\right) \rangle &=
    \left(\dfrac{2\ell+1}{2}\right)\dfrac{1}{(2\pi)^3N_{\mathrm{p}}(k)}\int_{V_{\bm{p}}}\int_{V_{\bm{q}}}\,d^3\bm{p}d^3\bm{q}
\, L_{\ell}\left(\mu\right)
\delta_\mathrm{D}\left(\bm{q}+\bm{p}\right)
\langle\delta_{g}^{\mathrm{s}}\left(\bm{q}\right)\delta_{g}^{\mathrm{s}}\left(\bm{p}\right)\rangle 
\notag \\
&= 
    \left(\dfrac{2\ell+1}{2}\right)\dfrac{1}{(2\pi)^3N_{\mathrm{p}}(k)}\int_{V_{\bm{p}}}\int_{V_{\bm{q}}}\,d^3\bm{p}d^3\bm{q}
\, L_{\ell}\left(\mu\right)
\delta_\mathrm{D}\left(\bm{q}+\bm{p}\right)^2 (2\pi)^3\mathrm{P}_{\mathrm{g}}^{\mathrm{s}}(\bm{p})
\notag \\
&= 
    \left(\dfrac{2\ell+1}{2}\right)\dfrac{1}{(2\pi)^3N_{\mathrm{p}}(k)}\int_{V_{\bm{p}}}\int_{V_{\bm{q}}}\,d^3\bm{p}d^3\bm{q}
\, L_{\ell}\left(\mu\right)
\delta_\mathrm{D}\left(\bm{q}+\bm{p}\right)V_{\mathrm{e}}\,\mathrm{P}_{\mathrm{g}}^{\mathrm{s}}(\bm{p})
\notag \\
&= 
    \left(\dfrac{2\ell+1}{2}\right)\dfrac{1}{V_{\mathrm{e}}V_k}\int_{V_{\bm{p}}}\int_{V_{\bm{q}}}\,d^3\bm{p}d^3\bm{q}
\, L_{\ell}\left(\mu\right)
\delta_\mathrm{D}\left(\bm{q}+\bm{p}\right)V_{\mathrm{e}}\,\mathrm{P}_{\mathrm{g}}^{\mathrm{s}}(\bm{p})
\notag \\
&= 
    \left(\dfrac{2\ell+1}{2}\right)\dfrac{1}{V_k}\int_{V_{\bm{p}}}\int_{V_{\bm{q}}}\,d^3\bm{p}d^3\bm{q}
\, L_{\ell}\left(\mu\right)
\delta_\mathrm{D}\left(\bm{q}+\bm{p}\right)\,\mathrm{P}_{\mathrm{g}}^{\mathrm{s}}(\bm{p}) 
\notag \\
&\approx \left(\dfrac{2\ell+ 1}{2}\right)\int^{+1}_{-1}d\mu\,\mathrm{P}_{\mathrm{g}}^{\mathrm{s}}\left(k, \mu\right)L_{\ell}\left(\mu\right),
\end{align}

\noindent where we used the approximation made in \cite{2009A&A...508.1193J} that $\delta_{\mathrm{D}}^2\approx\dfrac{V_{\mathrm{e}}}{(2\pi)^3}\delta_{\mathrm{D}} = k_f^{-3}\delta_{\mathrm{D}}$. In the last step it has been made the common approximation that $\bm{p}$ and $\bm{q}$ are very close to $k$ in module for thin enough shells (small $\Delta k$). The standard definition of the redshift galaxy power spectrum has also been used:

\begin{eqnarray}
\langle\delta_{g}^{\mathrm{s}}\left(\bm{q}\right)\delta_{g}^{\mathrm{s}}\left(\bm{p}\right)\rangle \,=\,(2\pi)^3 \delta_\mathrm{D}\left(\bm{q}+\bm{p}\right)\mathrm{P}_{\mathrm{g}}^{\mathrm{s}}(\bm{p}).
\end{eqnarray}

The redshift space galaxy bispectrum is defined as:

\begin{eqnarray}
\langle\delta_{g}^{\mathrm{s}}\left(\bm{q}_1\right)\delta_{g}^{\mathrm{s}}\left(\bm{q}_2\right)\delta_{g}^{\mathrm{s}}\left(\bm{q}_3\right)\rangle \,=\,(2\pi)^3 \delta_\mathrm{D}\left(\bm{q}_1+\bm{q}_2+\bm{q}_3\right)\mathrm{B}_{\mathrm{g}}^{\mathrm{s}}(\bm{q}_1,\bm{q}_2,\bm{q}_3).
\end{eqnarray}

\noindent The analytical expression for the bispectrum monopole model was given in Eq. \ref{bk_0_esp}.

\begin{figure*}
\centering
\includegraphics[width=0.7\textwidth]{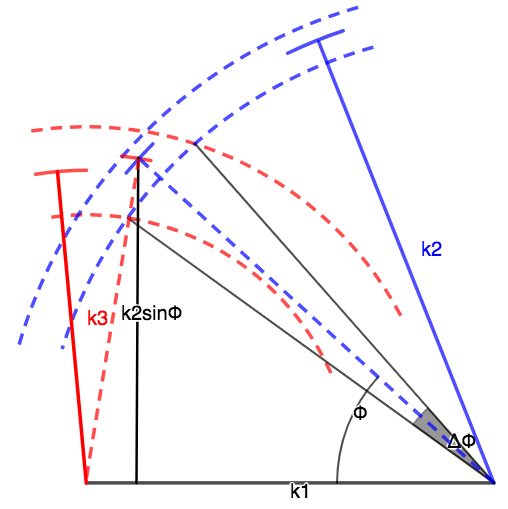}
\caption{Computation of the integration volume in Fourier space in the case of the bispectrum monopole. Once the side $k_1$ of the triangle is fixed, the other two sides are free to vary in the intersection given by two sphere of radius $k_2 - \Delta k_2/2\leq r_2 \leq k_2 + \Delta k_2/2$ and $k_3 - \Delta k_3/2\leq r_3 \leq k_3 + \Delta k_3/2$ respectively. In the Figure above the 2D projection of the annuli of thickness $\Delta k_2$ (blue) and $\Delta k_2$ (red) are shown. The angle $\phi$ correspond to the angle $\phi_{12}$ in the text.}
\label{fig:triangle_integration_volume}
\end{figure*}

\noindent Analogously to the power spectrum multipoles, the estimator for the bispectrum monopole can be defined as:

\begin{eqnarray}
\label{bk_0_est}
\hat{\mathrm{B}}^{(0)}_{\mathrm{g}}\left(k_1,k_2,k_3\right)
=
\dfrac{1}{4\pi}\dfrac{V_{\mathrm{e}}(2\pi)^{-6}}{N_{\mathrm{t}}(k_1,k_2,k_3)}\prod^3_{i=1}\int_{V_{\bm{q}_i}} \,d^3\bm{q}_i\,
\delta_{D}\left(\bm{q}_1+\bm{q}_2+\bm{q}_3\right)
\delta_{g}^{\mathrm{s}}\left(\bm{q}_1\right)\delta_{g}^{\mathrm{s}}\left(\bm{q}_2\right)\delta_{g}^{\mathrm{s}}\left(\bm{q}_3\right)
.
\end{eqnarray}

\noindent where $N_{\mathrm{t}}(k_1,k_2,k_3)$ is the number of independent grid points triplets inside the integration volume in Fourier space. As shown in the weak lensing 2D case by \citet{2013MNRAS.429..344K}, this is computed as:

\begin{eqnarray}
\label{eq:ntri}
N_{\mathrm{t}}(k_1,k_2,k_3) = \dfrac{V_{k_{123}}}{k_f^6} = k_f^{-6}\,\int_{V_{\bm{q}_1}}\int_{V_{\bm{q}_2}}\int_{V_{\bm{q}_3}}\,d^3\bm{q}_1d^3\bm{q}_2d^3\bm{q}_3\,
\delta_\mathrm{D}\,\left(\bm{q}_1+\bm{q}_2+\bm{q}_3\right) \simeq \dfrac{ 8\pi^2 k_1k_2k_3\Delta k_1\Delta k_2\Delta k_3}{k_f^6}.
\end{eqnarray}

\noindent It is important to notice that the result of the above integral must be symmetric in the $k$-vectors arguments. Therefore, the best way to derive the integral results is through geometrical considerations. Starting from $q_1$, this can be chosen in a spherical shell with volume $V_{k_1}\simeq 4\pi k_1^2\Delta k$. Once $q_1$ is fixed, considering the plane in which both $q_2$ and $q_3$ lie, they must connect to each other inside the 2D intersection formed by the two annuli defined by  $k_2-\Delta k_2/2 \leq q_2 \leq k_2+\Delta k_2/2 $  and  $k_3-\Delta k_3/2 \leq q_3 \leq k_3+\Delta k_3/2 $. This has approximately an area equal to $A_{k_{23}}\simeq k_2\Delta\phi_{12} \Delta k_2$. From Figure \ref{fig:triangle_integration_volume} it is possible to see that $\Delta \phi_{12}$ is defined by varying $k_3$ by $\Delta k_3$. $\phi_{12}$ can be obtained from:

\begin{eqnarray}
\cos \phi_{12} = \dfrac{k_1^2 +k_2^2 - k_3^2}{2k_1k_2},
\end{eqnarray}

\noindent
and therefore $\Delta\phi_{12}$ can be found differentiating with respect to $k_3$:

\begin{eqnarray}
\dfrac{d\cos\phi_{12}}{dk_3} = - \dfrac{d\phi_{12}}{dk_3}\sin\phi_{12} = -\dfrac{k_3}{k_1k_2} 
\qquad \Longrightarrow \qquad 
\Delta \phi_{12} = \dfrac{\Delta k_3 k_3}{k_1k_2}\left(\sin\phi_{12}\right)^{-1}
.
\end{eqnarray}
\noindent Finally the volume of the intersection between $k_2$ and $k_3$ is obtained by rotating the area just found around the axis defined by $k_1$:

\begin{eqnarray}
\label{eq:vk23}
V_{k_{23}} = 2\pi A_{k_{23}}  \left(k_2 \sin\phi_{12}\right)  ,
\end{eqnarray}

\noindent which allows to compute $V_{k_{123}} = V_{k_{1}} V_{k_{23}}$ in Eq. \ref{eq:ntri}.


\subsection{Power spectrum monopole and quadrupole covariance matrix: Gaussian term}
\label{sec:cov_pp}

\noindent Following the Appendix of \citet{2018MNRAS.tmp..252G} we can check that also the bispectrum monopole estimator defined in Eq. \ref{bk_0_est} is unbiased. Moreover it is possible to compute the Gaussian term of the covariance for the power spectrum monopole and quadrupole as follows:

\begin{align}
\label{eq:cov_pp}
&\mathrm{C}_\mathrm{G}^{\mathrm{P}^{(\ell)}_{\mathrm{g}}\mathrm{P}^{(\ell)}_{\mathrm{g}}}\left(k_1;k_2\right) = 
\left(\dfrac{2\ell+1}{2}\right)^2\dfrac{(2\pi)^{-6}}{N_{\mathrm{p}}\left(k_1\right)N_{\mathrm{p}}\left(k_2\right)}
\int_{V_{\bm{q}_1}}\int_{V_{\bm{q}_2}}\int_{V_{\bm{p}_1}}\int_{V_{\bm{p}_2}}
 d^3\bm{q}_1 d^3\bm{q}_2 d^3\bm{p}_1 d^3\bm{p}_2
L_{\ell}\left(\mu_1\right)L_{\ell}\left(\mu_2\right)
\delta_D\left(\bm{q}_1+\bm{p}_1\right)\delta_D\left(\bm{q}_2+\bm{p}_2\right) 
\notag \\
&\times
2(2\pi)^6\delta_D\left(\bm{q}_1+\bm{q}_2\right)\delta_D\left(\bm{p}_1+\bm{p}_2\right)
\mathrm{P}_{\mathrm{g}}^{\mathrm{s}}\left(\bm{q}_1\right)\mathrm{P}_{\mathrm{g}}^{\mathrm{s}}\left(\bm{p}_2\right)
\notag \\
&=
\left(\dfrac{2\ell+1}{2}\right)^2\dfrac{2}{N_{\mathrm{p}}\left(k_1\right)N_{\mathrm{p}}\left(k_2\right)}
\int_{V_{\bm{q}_1}}\int_{V_{\bm{q}_2}}
 d^3\bm{q}_1 d^3\bm{q}_2
L_{\ell}\left(\mu_1\right)L_{\ell}\left(\mu_2\right)
\delta_D\left(\bm{q}_1+\bm{q}_2\right)^2
\mathrm{P}_{\mathrm{g}}^{\mathrm{s}}\left(\bm{q}_1\right)\mathrm{P}_{\mathrm{g}}^{\mathrm{s}}\left(\bm{q}_2\right)
\notag \\
&=
\left(\dfrac{2\ell+1}{2}\right)^2\dfrac{2k_f^{-3}}{N_{\mathrm{p}}\left(k_1\right)N_{\mathrm{p}}\left(k_2\right)}
\int_{V_{\bm{q}_1}}\int_{V_{\bm{q}_2}}
 d^3\bm{q}_1 d^3\bm{q}_2
L_{\ell}\left(\mu_1\right)L_{\ell}\left(\mu_2\right)
\delta_D\left(\bm{q}_1+\bm{q}_2\right)
\mathrm{P}_{\mathrm{g}}^{\mathrm{s}}\left(\bm{q}_1\right)\mathrm{P}_{\mathrm{g}}^{\mathrm{s}}\left(\bm{q}_2\right)
\notag \\
&\approx 
\left(\dfrac{2\ell+1}{2}\right)^2\dfrac{2k_f^{-3}}{N_{\mathrm{p}}\left(k_1\right)N_{\mathrm{p}}\left(k_2\right)}\quad
\mathrm{P}_{\mathrm{g}}^{(\ell)}\left(k_1\right)
\mathrm{P}_{\mathrm{g}}^{(\ell)}\left(k_2\right)
\int_{V_{\bm{q}_1}}\int_{V_{\bm{q}_2}}d^3\bm{q}_1d^3\bm{q}_2\delta_D\left(\bm{q}_1+\bm{q}_2\right)
\notag \\
&=
\left(\dfrac{2\ell+1}{2}\right)^2\dfrac{2\delta^{\mathrm{K}}_{12}}{N_{\mathrm{p}}\left(k_1\right)}\quad\mathrm{P}_{\mathrm{g}}^{(\ell)}\left(k_1\right)^2 ,
\end{align}

\noindent where again we used the approximation made in \cite{2009A&A...508.1193J} that $\delta_{\mathrm{D}}^2\approx\dfrac{V_{\mathrm{e}}}{(2\pi)^3}\delta_{\mathrm{D}} = k_f^{-3}\delta_{\mathrm{D}}$. $\delta^{\mathrm{K}}_{12}$ is the Kronecker delta indicating that the vector $\bm{q}_1$ and $\bm{q}_2$ are identical (in the second step trivial $\delta_{\mathrm{K}}$ have been omitted in order to avoid making the notation heavier by adding also the wave-vector letter). In the last steps we made the approximation that the power spectrum monopole and quadrupoles do not vary significantly when integrated over the bin in Fourier space.


\subsection{Bispectrum monopole covariance matrix: Gaussian term}
\label{sec:cov_bb_g}
Analogously to the above we now compute the diagonal term of the bispectrum monopole covariance matrix:

\begin{align}
\label{eq:cov_bb_g}
&\mathrm{C}_\mathrm{G}^{\mathrm{B}^0_{\mathrm{g}}\mathrm{B}^0_{\mathrm{g}}}\left(k_1,k_2,k_3;k_4,k_5,k_6\right) = 
\notag \\
&=
\dfrac{1}{16\pi^2}\dfrac{(2\pi k_f)^{-6}}{N_{\mathrm{t}}\left(k_1,k_2,k_3\right)N_{\mathrm{t}}\left(k_4,k_5,k_6\right)}
\prod^6_{i=1}\int_{V_{\bm{q}_i}} d^3\bm{q}_i
\delta_D\left(\bm{q}_1+\bm{q}_2+\bm{q}_3\right)\delta_D\left(\bm{q}_4+\bm{q}_5+\bm{q}_6\right)
\notag\\
&\times
(2\pi)^9\delta_D\left(\bm{q}_1+\bm{q}_4\right)\delta_D\left(\bm{q}_2+\bm{q}_5\right)\delta_D\left(\bm{q}_3+\bm{q}_6\right)
\mathrm{P}_{\mathrm{g}}^{\mathrm{s}}\left(\bm{q}_1\right)\mathrm{P}_{\mathrm{g}}^{\mathrm{s}}\left(\bm{p}_2\right)\mathrm{P}_{\mathrm{g}}^{\mathrm{s}}\left(\bm{q}_3\right) \quad + 5 \quad\mathrm{perm.}
\notag \\
&=
\dfrac{\mathrm{D}_{123456}}{16\pi^2}\dfrac{(2\pi)^3 k_f^{-6}}{N_{\mathrm{t}}\left(k_1,k_2,k_3\right)^2}
\prod^3_{i=1}\int_{V_{\bm{q}_i}} d^3\bm{q}_i
\delta_D\left(\bm{q}_1+\bm{q}_2+\bm{q}_3\right)^2
\mathrm{P}_{\mathrm{g}}^{\mathrm{s}}\left(\bm{q}_1\right)\mathrm{P}_{\mathrm{g}}^{\mathrm{s}}\left(\bm{p}_2\right)\mathrm{P}_{\mathrm{g}}^{\mathrm{s}}\left(\bm{q}_3\right)
\notag \\
&=
\dfrac{\mathrm{D}_{123456}}{16\pi^2}\dfrac{V_{\mathrm{e}} k_f^{-6}}{N_{\mathrm{t}}\left(k_1,k_2,k_3\right)^2}
\prod^3_{i=1}\int_{V_{\bm{q}_i}} d^3\bm{q}_i
\delta_D\left(\bm{q}_1+\bm{q}_2+\bm{q}_3\right)
\mathrm{P}_{\mathrm{g}}^{\mathrm{s}}\left(\bm{q}_1\right)\mathrm{P}_{\mathrm{g}}^{\mathrm{s}}\left(\bm{p}_2\right)\mathrm{P}_{\mathrm{g}}^{\mathrm{s}}\left(\bm{q}_3\right)
\notag \\
&\approx
\dfrac{\mathrm{D}_{123456}}{16\pi^2}\dfrac{V_{\mathrm{e}} k_f^{-6}}{N_{\mathrm{t}}\left(k_1,k_2,k_3\right)^2}
\mathrm{P}_{\mathrm{g}}^{(0)}\left(k_1\right)\mathrm{P}_{\mathrm{g}}^{(0)}\left(k_2\right)\mathrm{P}_{\mathrm{g}}^{(0)}\left(k_3\right)
\prod^3_{i=1}\int_{V_{\bm{q}_i}} d^3\bm{q}_i
\delta_D\left(\bm{q}_1+\bm{q}_2+\bm{q}_3\right)
\notag\\
&=
\dfrac{\mathrm{D}_{123456}}{16\pi^2}\dfrac{V_{\mathrm{e}}}{N_{\mathrm{t}}\left(k_1,k_2,k_3\right)}
\mathrm{P}_{\mathrm{g}}^{(0)}\left(k_1\right)\mathrm{P}_{\mathrm{g}}^{(0)}\left(k_2\right)\mathrm{P}_{\mathrm{g}}^{(0)}\left(k_3\right)
,
\end{align}

\noindent where $\mathrm{D}_{123456}$ stands for all the possible permutations and has values $6,2,1$ respectively for equilateral, isosceles and scalene triangles. Again it has been assumed that the power spectrum monopole does not vary significantly inside the integration volume.


\subsection{Bispectrum monopole covariance matrix: non-Gaussian term}
\label{sec:cov_bb_ng}
In this work we use only one of the non-Gaussian terms of the bispectrum monopole covariance matrix. This is because we just need to model the covariance matrix analytically in order to derive the weights for the compression. This additional term allows to better capture the correlation between different triangle bins. We leave to future work the analytic computation of the remaining terms.

\begin{align}
\label{eq:cov_bb_ng}
&\mathrm{C}_\mathrm{NG}^{\mathrm{B}^0_{\mathrm{g}}\mathrm{B}^0_{\mathrm{g}}}\left(k_1,k_2,k_3;k_4,k_5,k_6\right) = 
\notag \\
&=
\dfrac{1}{16\pi^2}\dfrac{(2\pi k_f)^{-6}}{N_{\mathrm{t}}\left(k_1,k_2,k_3\right)N_{\mathrm{t}}\left(k_4,k_5,k_6\right)}
\prod^6_{i=1}\int_{V_{\bm{q}_i}} d^3\bm{q}_i
\delta_D\left(\bm{q}_1+\bm{q}_2+\bm{q}_3\right)\delta_D\left(\bm{q}_4+\bm{q}_5+\bm{q}_6\right)
\notag\\
&\times
(2\pi)^6\delta_D\left(\bm{q}_1+\bm{q}_2+\bm{q}_4\right)\delta_D\left(\bm{q}_3+\bm{q}_5+\bm{q}_6\right)
\mathrm{B}_{\mathrm{g}}^{\mathrm{s}}\left(\bm{q}_1,\bm{q}_2,\bm{q}_4\right)\mathrm{B}_{\mathrm{g}}^{\mathrm{s}}\left(\bm{q}_3,\bm{q}_5,\bm{q}_6\right) \quad + 8 \quad\mathrm{perm.}
\notag \\
&=
\dfrac{1}{16\pi^2}\dfrac{ k_f^{-6}\delta^{\mathrm{K}}_{34}}{N_{\mathrm{t}}\left(k_1,k_2,k_3\right)N_{\mathrm{t}}\left(k_3,k_5,k_6\right)}
\int_{V_{\bm{q}_1}}\int_{V_{\bm{q}_2}}\int_{V_{\bm{q}_3}}\int_{V_{\bm{q}_5}}\int_{V_{\bm{q}_6}} d^3\bm{q}_1d^3\bm{q}_2d^3\bm{q}_3d^3\bm{q}_5d^3\bm{q}_6
\,\delta_D\left(\bm{q}_1+\bm{q}_2+\bm{q}_3\right)
\notag\\
&\times
\delta_D\left(\bm{q}_3+\bm{q}_5+\bm{q}_6\right)^2
\mathrm{B}_{\mathrm{g}}^{\mathrm{s}}\left(\bm{q}_1,\bm{q}_2,-\bm{q}_3\right)\mathrm{B}_{\mathrm{g}}^{\mathrm{s}}\left(\bm{q}_3,\bm{q}_5,\bm{q}_6\right) \quad + 8 \quad\mathrm{perm.}
\notag \\
&=
\dfrac{1}{16\pi^2}\dfrac{ k_f^{-9}\delta^{\mathrm{K}}_{34}}{N_{\mathrm{t}}\left(k_1,k_2,k_3\right)N_{\mathrm{t}}\left(k_3,k_5,k_6\right)}
\int_{V_{\bm{q}_1}}\int_{V_{\bm{q}_2}}\int_{V_{\bm{q}_3}}\int_{V_{\bm{q}_5}}\int_{V_{\bm{q}_6}}
d^3\bm{q}_1d^3\bm{q}_2d^3\bm{q}_3d^3\bm{q}_5d^3\bm{q}_6 \,
\delta_D\left(\bm{q}_1+\bm{q}_2+\bm{q}_3\right)
\notag\\
&\times
\delta_D\left(\bm{q}_3+\bm{q}_5+\bm{q}_6\right)
\mathrm{B}_{\mathrm{g}}^{\mathrm{s}}\left(\bm{q}_1,\bm{q}_2,-\bm{q}_3\right)\mathrm{B}_{\mathrm{g}}^{\mathrm{s}}\left(\bm{q}_3,\bm{q}_5,\bm{q}_6\right) \quad + 8 \quad\mathrm{perm.}
\notag \\
&\approx
\dfrac{1}{16\pi^2}\dfrac{ k_f^{-3}\delta^{\mathrm{K}}_{34}}{N_{\mathrm{t}}\left(k_3,k_5,k_6\right)}
\mathrm{B}_{\mathrm{g}}^{(0)}\left(k_1,k_2,k_3\right)\mathrm{B}_{\mathrm{g}}^{(0)}\left(k_3,k_5,k_6\right)
\int_{V_{\bm{q}_i}} d^3\bm{q}_5d^3\bm{q}_6\,
\delta_D\left(\bm{q}_3+\bm{q}_5+\bm{q}_6\right)
 \quad + 8 \quad\mathrm{perm.}
\notag \\
&=
\dfrac{\delta^{\mathrm{K}}_{34}}{16\pi^2}\dfrac{k_f^{3}}{4\pi k_3^2\Delta k_3}
\mathrm{B}_{\mathrm{g}}^{(0)}\left(k_1,k_2,k_3\right)\mathrm{B}_{\mathrm{g}}^{(0)}\left(k_3,k_5,k_6\right)
\quad + 8 \quad\mathrm{perm.}
,
\end{align}

\noindent where the usual approximations have been used together with Eq. \ref{eq:vk23} which in the last step has been used to simplify the integration over the volume in Fourier space once one of the $\bm{k}$-vectors is fixed.


\subsection{Cross-covariance term}
\label{sec:cov_cross}
For what concerns the cross-covariance term between power spectrum (monopole/quadrupole) and bispectrum monopole we use only the first leading term in our model:

\begin{align}
\label{eq:cov_cross}
&\mathrm{C}^{\mathrm{P}^{(\ell)}_{\mathrm{g}}\mathrm{B}^0_{\mathrm{g}}}\left(k_1;k_2,k_3,k_4\right) = 
\notag \\
&=
\dfrac{1}{4\pi}\left(\dfrac{2\ell+1}{2}\right)\dfrac{(2\pi)^{-6}k_f^{-3}}{N_{\mathrm{p}}\left(k_1\right)N_{\mathrm{t}}\left(k_2,k_3,k_4\right)}
\int_{V_{\bm{q}_1}}\int_{V_{\bm{p}_1}} d^3\bm{q}_1d^3\bm{p}_1\prod^4_{i=2}\int_{V_{\bm{q}_i}} d^3\bm{q}_i
\delta_D\left(\bm{q}_1+\bm{p}_1\right)\delta_D\left(\bm{q}_2+\bm{q}_3+\bm{q}_4\right)\,L_{\ell}\left(\mu_1\right)
\notag\\
&\times
2(2\pi)^6\delta_D\left(\bm{q}_1+\bm{q}_2\right)\delta_D\left(\bm{p}_1+\bm{q}_3+\bm{q}_4\right)
\mathrm{P}_{\mathrm{g}}^{\mathrm{s}}\left(\bm{q}_2\right)\mathrm{B}_{\mathrm{g}}^{\mathrm{s}}\left(\bm{q}_2,\bm{q}_3,\bm{q}_4\right) \quad + 2 \quad\mathrm{perm.}
\notag \\
&=
\dfrac{1}{2\pi}\left(\dfrac{2\ell+1}{2}\right)\dfrac{k_f^{-3}}{N_{\mathrm{p}}\left(k_1\right)N_{\mathrm{t}}\left(k_2,k_3,k_4\right)}
\prod^4_{i=1}\int_{V_{\bm{q}_i}} d^3\bm{q}_i
\,L_{\ell}\left(\mu_1\right)
\delta_D\left(\bm{q}_1+\bm{q}_2\right)\delta_D\left(\bm{q}_2+\bm{q}_3+\bm{q}_4\right)^2
\mathrm{P}_{\mathrm{g}}^{\mathrm{s}}\left(\bm{q}_2\right)\mathrm{B}_{\mathrm{g}}^{\mathrm{s}}\left(\bm{q}_2,\bm{q}_3,\bm{q}_4\right) \quad + 2 \quad\mathrm{perm.}
\notag \\
&=
\dfrac{1}{2\pi}\left(\dfrac{2\ell+1}{2}\right)\dfrac{k_f^{-6}\delta^{\mathrm{K}}_{12}}{N_{\mathrm{p}}\left(k_2\right)N_{\mathrm{t}}\left(k_2,k_3,k_4\right)}
\prod^4_{i=2}\int_{V_{\bm{q}_i}} d^3\bm{q}_i
\,L_{\ell}\left(\mu_2\right) \delta_D\left(\bm{q}_2+\bm{q}_3+\bm{q}_4\right)
\mathrm{P}_{\mathrm{g}}^{\mathrm{s}}\left(\bm{q}_2\right)\mathrm{B}_{\mathrm{g}}^{\mathrm{s}}\left(\bm{q}_2,\bm{q}_3,\bm{q}_4\right) \quad + 2 \quad\mathrm{perm.}
\notag \\
& \approx
\dfrac{1}{2\pi}\left(\dfrac{2\ell+1}{2}\right)\dfrac{\delta^{\mathrm{K}}_{12}}{N_{\mathrm{p}}\left(k_2\right)}
\mathrm{P}^{(\ell)}_{\mathrm{g}}\left(k_2\right)\mathrm{B}^{(0)}_{\mathrm{g}}\left(k_2,k_3,k_4\right) \quad + 2 \quad\mathrm{perm.}
,
\end{align}

\noindent where once more we have used the same approximation of the power spectrum multipoles and bispectrum monopole not varying significantly inside the integration volume.
\newpage

\section{Validation tests}
\label{sec:validation}

In Table \ref{tab:4pars_shifted_cosmology} we report the results obtained compressing the bispectrum with respect to the shifted fiducial parameter sets. This is to test whether the performance of the compression is affected by the choice of fiducial set of parameter values. In particular, we consider two cases by varying the fiducial cosmology by adding/subtracting $1\sigma$ 1D credible intervals (derived from the MCMC) to all the parameters. The table quantifies that the shifts in the means of the 1D posterior distributions produced by considering a non-optimal fiducial cosmology are small compared to the $1\sigma$ 1D credible intervals of the MCMC results.

\begin{figure}
    \centering
    \subfloat[ MC-KL]{{\includegraphics[width=0.5\textwidth]
    {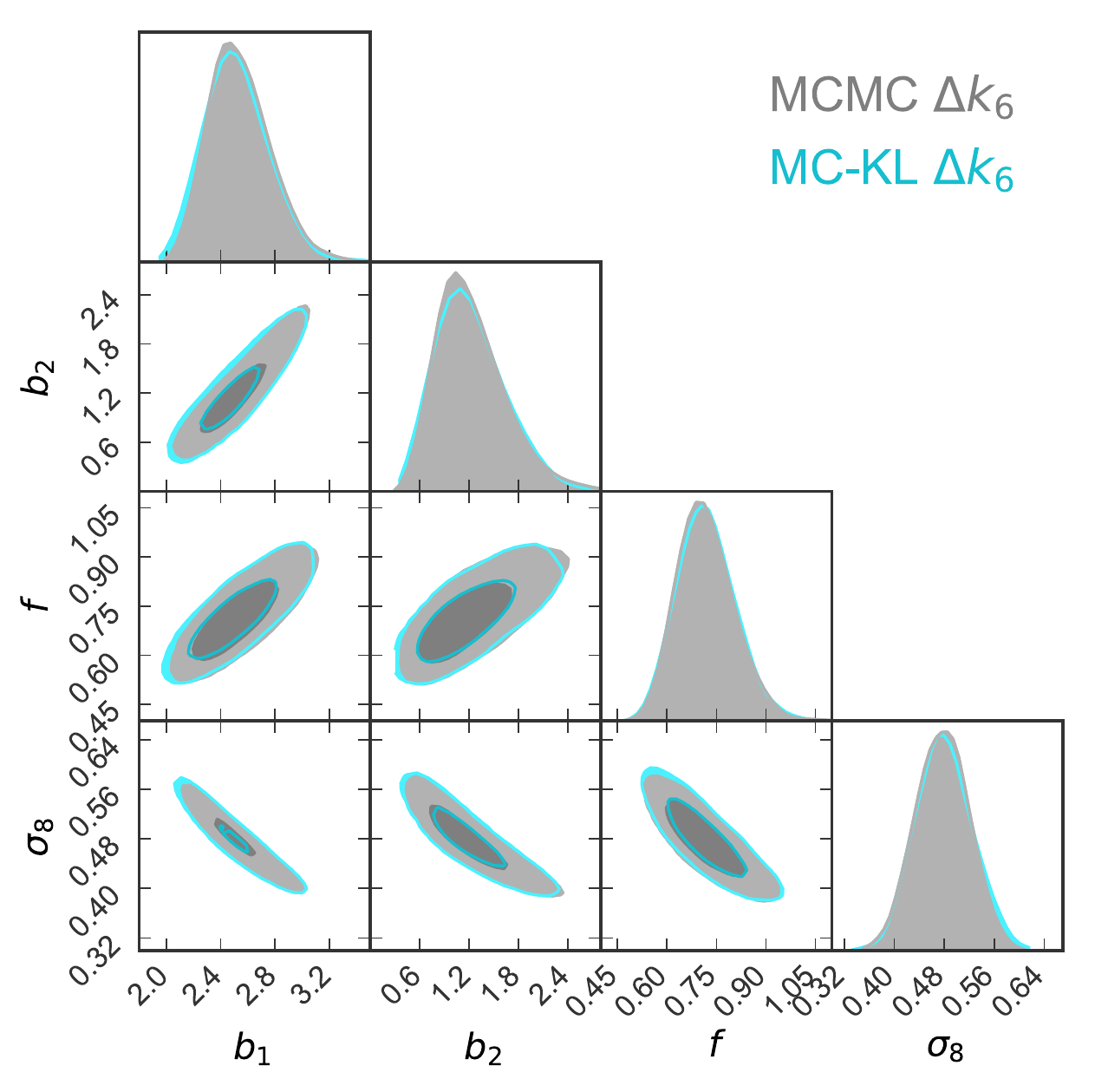}
    }}%
    \subfloat[G-PCA]{{\includegraphics[width=0.5\textwidth]
    {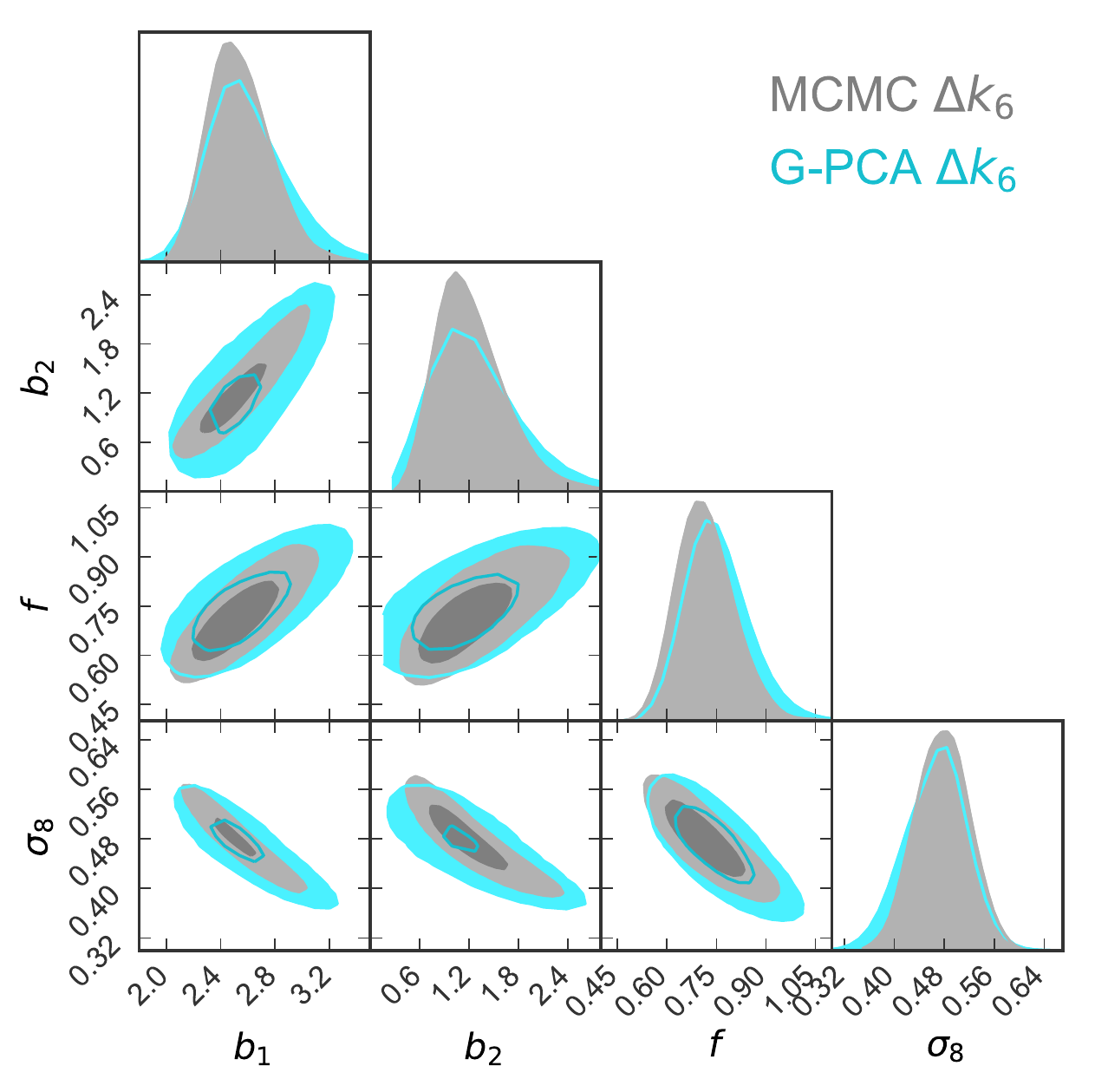}
    }}%
    \caption{
     Joint data-vector $\left[\mathrm{P}^{(0)}_{\mathrm{g}},\mathrm{P}^{(2)}_{\mathrm{g}},\mathrm{B}^{(0)}_{\mathrm{g}}\right]$ posteriors: MC-KL and G-PCA four-parameter $\Delta k_6$ case. \newline
    \textbf{a)} 
    2-D $68\%$ and $95\%$ credible regions are shown in order to compare the MC-KL (cyan) performance to the one of the standard MCMC (grey) for the full data  vector. The difference between MC-KL and MCMC contours is quantified in Table \ref{tab:4pars_consistency}.
    \newline
    \textbf{b)} The same as a) but for the G-PCA method.
    }
    \label{fig:mcmc_vs_MC-KL_vs_pc_check}
\end{figure}

\begin{figure}
    \centering
    \subfloat[ MC-KL]{{\includegraphics[width=0.5\textwidth]
    {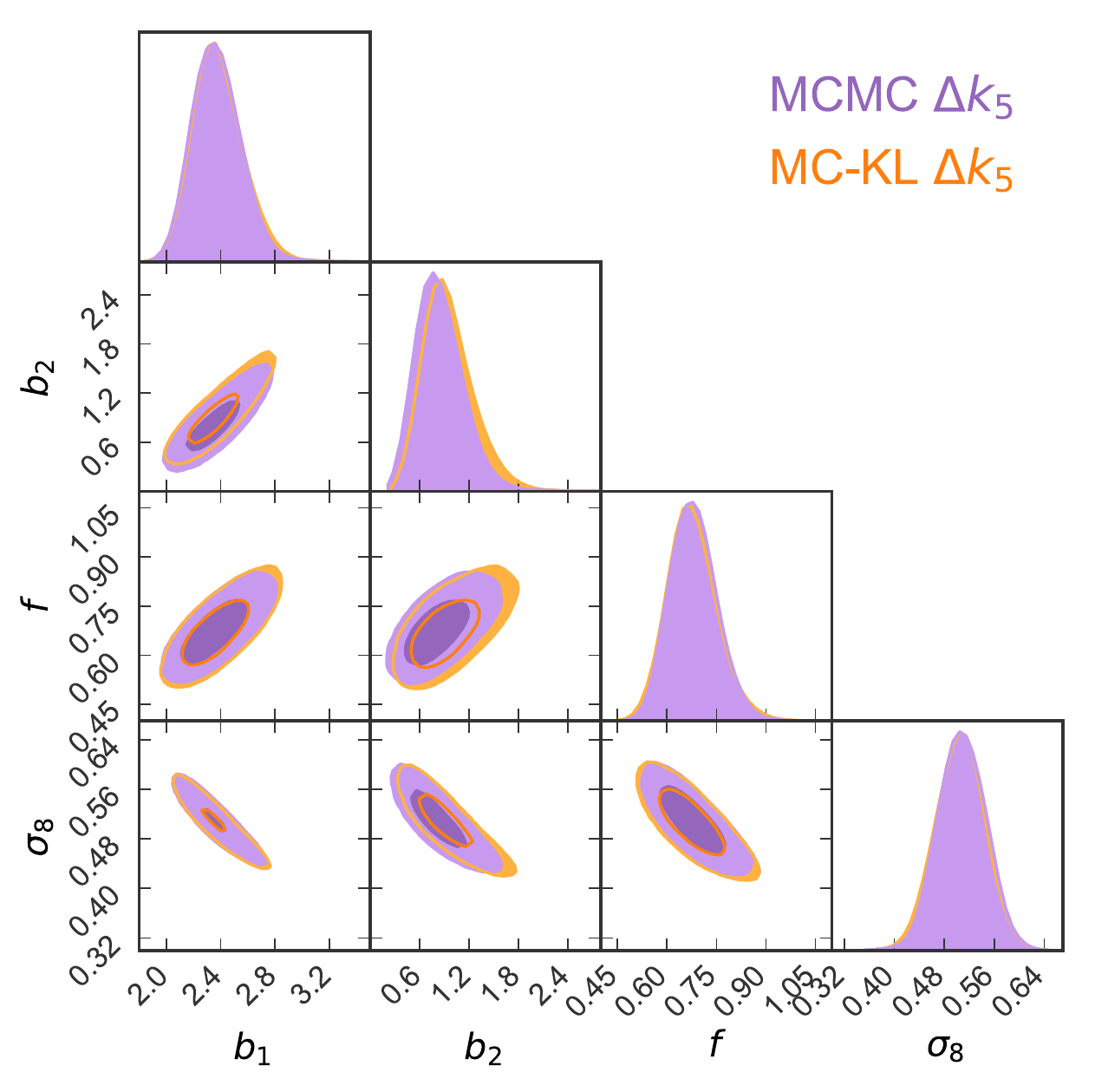}
    }}%
    \subfloat[G-PCA]{{\includegraphics[width=0.5\textwidth]
    {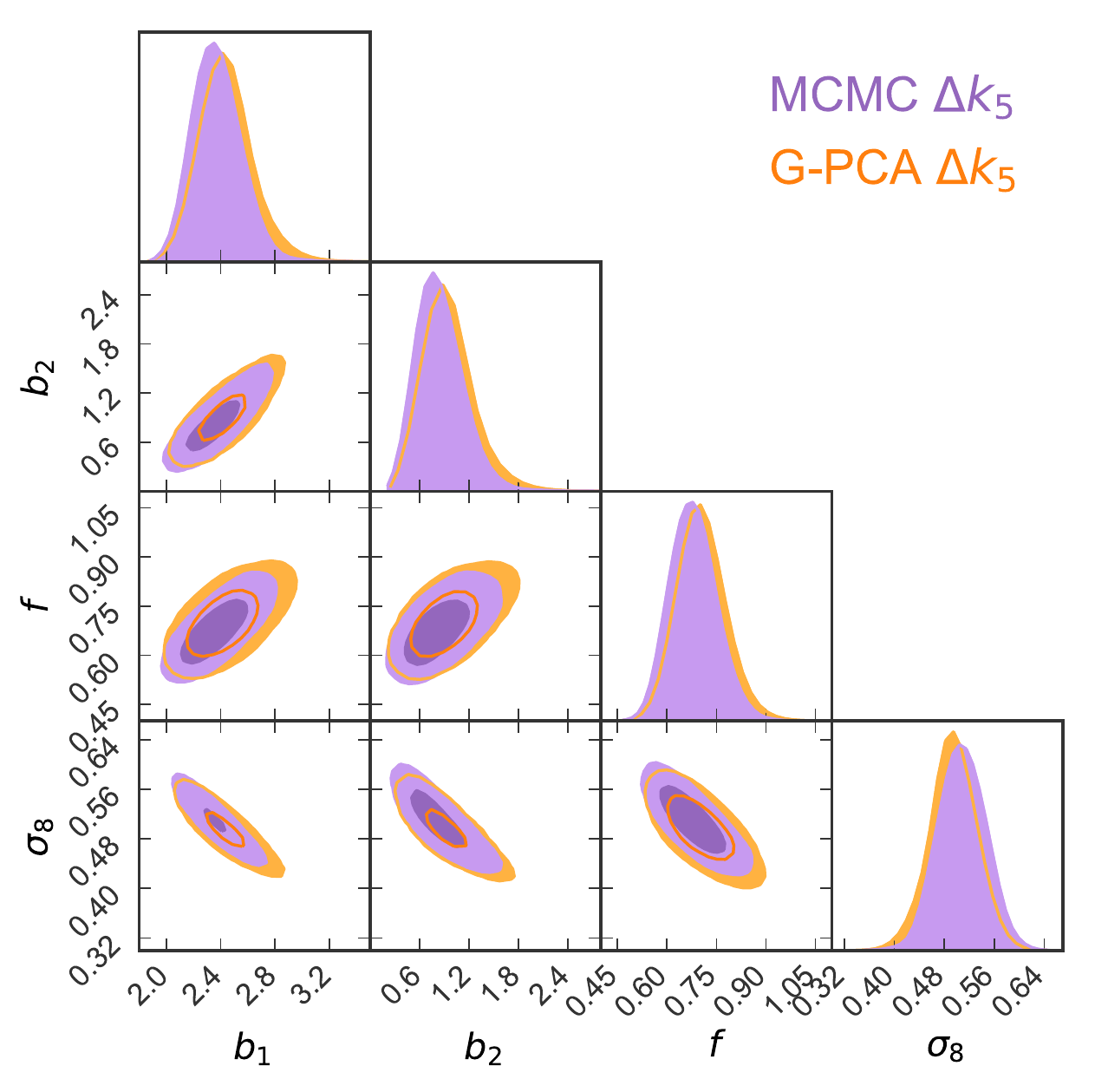}
    }}%
    \caption{
     Joint data-vector $\left[\mathrm{P}^{(0)}_{\mathrm{g}},\mathrm{P}^{(2)}_{\mathrm{g}},\mathrm{B}^{(0)}_{\mathrm{g}}\right]$ posteriors: MC-KL and G-PCA four-parameter $\Delta k_5$ case.\newline 
     Both \textbf{a)} and \textbf{b)} are the same as for Figure \ref{fig:mcmc_vs_MC-KL_vs_pc_check} for the $\Delta k_5$ case. 
    }
    \label{fig:mcmc_vs_MC-KL_vs_pc_check2}
\end{figure}

In Figures \ref{fig:mcmc_vs_MC-KL_vs_pc_check} and \ref{fig:mcmc_vs_MC-KL_vs_pc_check2} the 1 and 2-D posterior distributions obtained via MCMC/MC-KL/G-PCA for the test cases relative to the $\Delta k_6$ and $\Delta k_5$ binning cases are shown. 
MC-KL recovers with very good approximation the 1 and 2-D posterior distributions derived by the MCMC. G-PCA shows a slightly greater loss of information for the $\Delta k_6$ case. However this is noticeably closer to the MCMC/MC-KL result when the number of triangle bins used is increased ($\Delta k_5$ case).

\begin{figure}
\centering
\includegraphics[width=\textwidth]{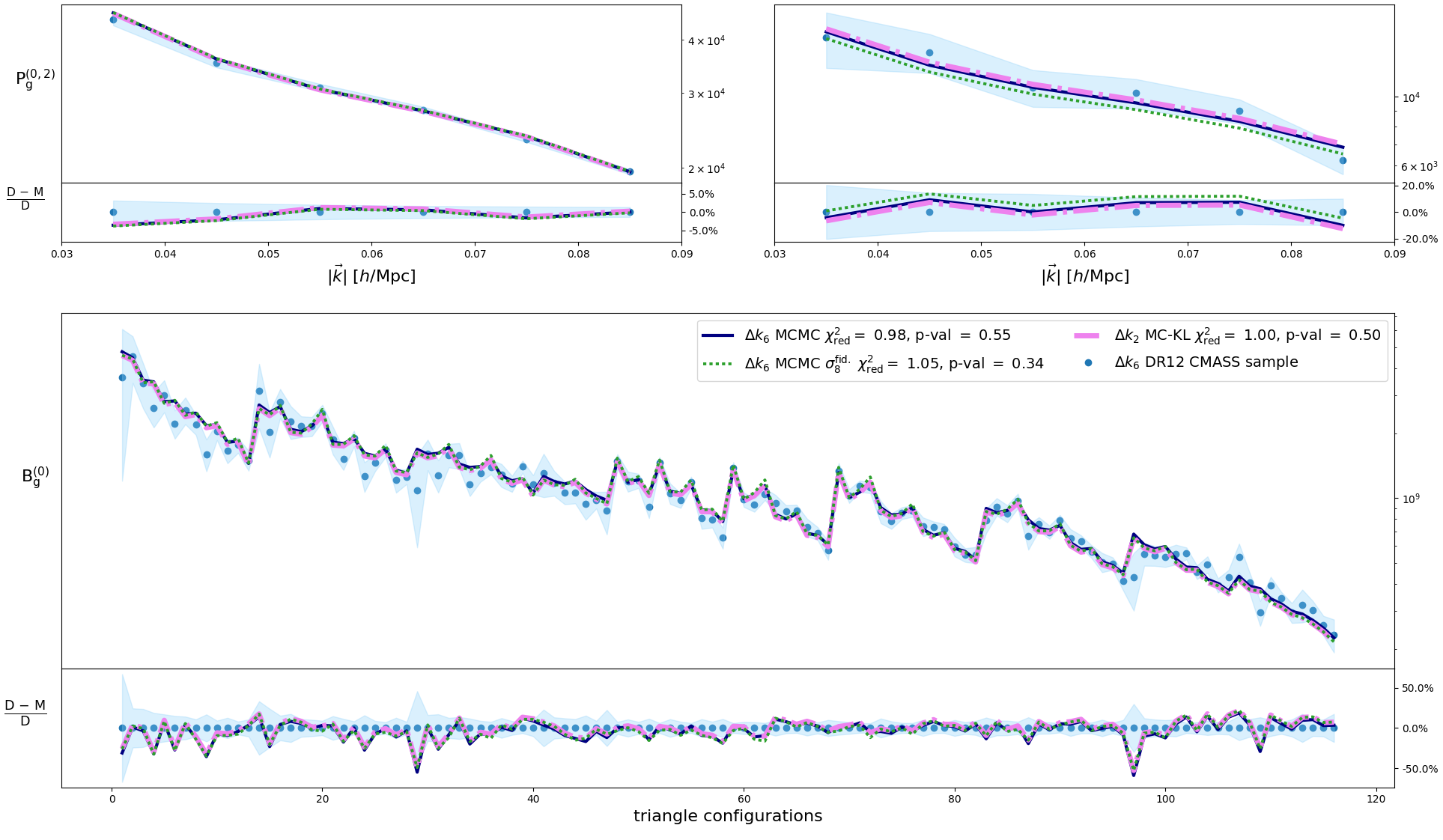}
\caption{Reduced $\chi^2$ and $p$-values for the best-fit parameters obtained using the MCMC/MC-KL methods with varying $\sigma_8$ and for the MCMC leaving $\sigma_8=\sigma_8^{\mathrm{fid.}}$ fixed . The $k$-binnings shown for the four parameter case ($b_1,b_2,f,\sigma_8$) are respectively the standard $\Delta k_6$ (navy) for the MCMC and the $\Delta k_2$ (pink) for the MC+KL. The line corresponding to the fit obtained by letting free to vary only the parameters ($b_1,b_2,f$) is shown in green.   The two upper panels are for the power spectrum monopole (left) and quadrupole (right) while the bottom panel refers to the bispectrum monopole. 
The lower part of each panel shows the relative difference between the data measurements and the different models.
Even if for example $b_1$ and $\sigma_8$ values are shifted in the cases of $\Delta k_6$ and $\Delta k_2$, this is due to the strong degeneracy between them and both models are practically identical to the one given by the three parameters fit ($b_1,b_2,f$) with $\sigma_8=\sigma_8^{\mathrm{fid.}}$. The only way to converge to the results obtained by the BOSS collaboration is to consider a larger range of scales (as they have done) for both power spectrum and bispectrum which however involves a more complex modelling of the data-vector.}
\label{fig:fixed_s8_chi2}
\end{figure}

In Figure \ref{fig:fixed_s8_chi2} we compare the best-fit model obtained by varying four parameters ($b_1,b_2,f,\sigma_8$) with the best-fit model corresponding to a fit done via standard MCMC sampling with only three parameters varied, ($b_1,b_2,f$), with $\sigma_8=\sigma_8^{\mathrm{fid.}}$. For the three parameter case we find running the MCMC: 
$b_1=1.98\pm0.01,b_2=0.39\pm0.06,$   $f(z_{\mathrm{CMASS}})=0.52\pm0.03$ with $\sigma_8^{\mathrm{fid.}}(z_{\mathrm{CMASS}})=0.61$.

Thereby we show that the discrepancy between the results of this paper and the ones presented in the BOSS collaboration analysis \citet{2017MNRAS.465.1757G} is, together with the different model used for the power spectrum monopole and quadrupole, probably due to the different range of scales considered. Indeed, by limiting our analysis to a smaller range of scales in $k$-space, the degeneracy between the amplitude-like parameters $b_1$ and $\sigma_8$ is much stronger. That is visible in Figure \ref{fig:fixed_s8_chi2}, where the models given by sets of parameters with very different $b_1$, $b_2$ and $\sigma_8$ parameters produce very similar predictions of the signals all with good $\chi^2_{\mathrm{red.}}$ and $p$-values.

\begin{table}
    \caption{ Four parameter-case, checking consistency for shifted fiducial cosmology. \newline
  \textbf{Upper half:} Mean values of the posterior distributions and $68\%$ credible intervals for the MCMC and the MC-KL / G-PCA compression methods. We report the values for the $\Delta k_6$ binning case for both compression methods in three cases consisting in using for the compression: the fiducial cosmology, the fiducial cosmology shifted by $+1\,\sigma$ and the fiducial cosmology shifted by $-1\,\sigma$.
  \newline
  \textbf{Lower half:} In the compression columns we report the relative difference between the posterior modes obtained via MCMC and the ones obtained via compression (MC-KL or G-PCA). In the MCMC columns the relative size of the $68\%$ credible intervals obtained via MCMC sampling is shown. By comparing the MCMC columns to the compression ones, it is clear that the difference between the mean parameter values obtained via MCMC and the ones obtained via compression (MC-KL or G-PCA) are evidently within the $68\%$ credible intervals given by the MCMC on the full data-vector.
  }
\makebox[\textwidth][c]{
    \begin{tabular}{cccccccc}
\toprule
        & \multicolumn{3}{c}{$\Delta k_6$} & \multicolumn{2}{c}{$\Delta k_6 \quad+\,1\sigma$}& \multicolumn{2}{c}{$\Delta k_6\quad-\,1\sigma$}  \\
\cmidrule(lr){2-4}\cmidrule(lr){5-6}\cmidrule(lr){7-8}
        &  MCMC  & MC-KL  & G-PCA  & MC-KL  & G-PCA    & MC-KL   & G-PCA  \\
\cmidrule(lr){2-4}\cmidrule(lr){5-6}\cmidrule(lr){7-8}
$b_1$      & 2.41 $\pm$ 0.22 & 2.41 $\pm$ 0.23 & 2.49 $\pm$ 0.27 & 2.47 $\pm$ 0.23 & 2.41 $\pm$ 0.12 & 2.54 $\pm$ 0.24 & 2.34 $\pm$ 0.37 \\
$b_2$      & 1.00 $\pm$ 0.40 & 1.04 $\pm$ 0.42 & 1.08 $\pm$ 0.47 & 1.04 $\pm$ 0.40 & 1.29 $\pm$ 0.25 & 1.03 $\pm$ 0.44 & 0.93 $\pm$ 0.67 \\
$f$        & 0.69 $\pm$ 0.08 & 0.72 $\pm$ 0.09 & 0.72 $\pm$ 0.09 & 0.70 $\pm$ 0.08 & 0.69 $\pm$ 0.05 & 0.72 $\pm$ 0.09 & 0.68 $\pm$ 0.12 \\
$\sigma_8$ & 0.50 $\pm$ 0.04 & 0.48 $\pm$ 0.05 & 0.48 $\pm$ 0.05 & 0.49 $\pm$ 0.04 & 0.49 $\pm$ 0.03 & 0.46 $\pm$ 0.05 & 0.50 $\pm$ 0.07 \\
\cmidrule(lr){2-4}\cmidrule(lr){5-6}\cmidrule(lr){7-8}
& $\dfrac{\Delta\theta^{\mathrm{mc}}}{\theta^{\mathrm{mc}}}\;\left[\%\right]$ 
& \multicolumn{2}{c}{$\dfrac{\theta^{\mathrm{comp.}} - \theta^{\mathrm{mc}}}{\theta^{\mathrm{mc}}}\;\left[\%\right]$}
& \multicolumn{2}{c}{$\dfrac{\theta^{\mathrm{comp.}} - \theta^{\mathrm{mc}}}{\theta^{\mathrm{mc}}}\;\left[\%\right]$}
& \multicolumn{2}{c}{$\dfrac{\theta^{\mathrm{comp.}} - \theta^{\mathrm{mc}}}{\theta^{\mathrm{mc}}}\;\left[\%\right]$} \\
\cmidrule(lr){2-2}\cmidrule(lr){3-4}\cmidrule(lr){5-6}\cmidrule(lr){7-8}
$ b_1 $      & 9.2  & -0.3 & 3.3  & 2.15  & -0.26 & 8.57  & 0.31 \\
$ b_2 $      & 40.3 & 3.5  & 7.5  & 3.47  & 28.68 & 25.29 & 13.26 \\
$ f   $      & 12.1 & 4.4  & 4.4  & 0.84  & 0.51  & 6.96  & 0.26  \\
$ \sigma_8$  & 8.5  & -5.1 & -5.5 & -3.25 & -2.91 & -8.94 & -1.39 \\
\bottomrule
\end{tabular}%
}
\label{tab:4pars_shifted_cosmology}%
\end{table}%

\label{lastpage}

\end{document}